\documentclass[review=false]{acmart}

\def\markup{0}

\if\markup1
\usepackage[normalem]{ulem} %1
\newcommand{\rv}[1]{{\leavevmode\color{blue}#1}}

\else
\newcommand{\rv}[1]{#1}
\newcommand{\sout}[1]{}
\fi

\DeclareUnicodeCharacter{FF09}{)}
\usepackage{graphicx}
\usepackage{caption}
\usepackage{subcaption}
\usepackage{multirow} 
\usepackage{multicol} 
\usepackage{booktabs}
\graphicspath{{Figures/}}
\usepackage{svg}
\usepackage{soul, color, xcolor}
\usepackage{tabularx,ragged2e,booktabs}

\usepackage{dblfloatfix} % 修复双栏布局中的浮动问题

\AtBeginDocument{%
  }

\copyrightyear{2025}
\acmYear{2025}
\setcopyright{acmlicensed}\acmConference[CHI '25]{CHI Conference on Human
Factors in Computing Systems}{April 26-May 1, 2025}{Yokohama, Japan}
\acmBooktitle{CHI Conference on Human Factors in Computing Systems (CHI
'25), April 26-May 1, 2025, Yokohama, Japan}
\acmDOI{10.1145/3706598.3713409}
\acmISBN{979-8-4007-1394-1/25/04}

\begin{document}

\title{Facilitating Daily Practice in Intangible Cultural Heritage through Virtual Reality: A Case Study of Traditional Chinese Flower Arrangement}

\author{Yingna Wang}
\affiliation{%
  \institution{The Hong Kong University of Science and Technology (Guangzhou)}
  \city{Guangzhou}
  \country{China}
}
\email{ywang885@connect.hkust-gz.edu.cn}

\author{Qingqin Liu}
\affiliation{%
  \institution{City University of Hong Kong}
  \city{Hong Kong SAR}
  \country{China}}
\email{neoliu-c@my.cityu.edu.hk}

\author{Xiaoying Wei}
\affiliation{%
  \institution{The Hong Kong University of Science and Technology}
  \city{Hong Kong SAR}
  \country{China}
}
\email{xweias@connect.ust.hk}

\author{Mingming Fan}
\authornote{Corresponding author}
\affiliation{%
  \institution{The Hong Kong University of Science and Technology (Guangzhou)}
  \city{Guangzhou}
  \country{China}
}
\affiliation{%
  \institution{The Hong Kong University of Science and Technology}
  \city{Hong Kong SAR}
  \country{China}}
\email{mingmingfan@ust.hk}



\renewcommand{\shorttitle}{Facilitating Daily Practice in Intangible Cultural Heritage through Virtual Reality}

\begin{abstract}

The essence of intangible cultural heritage (ICH) lies in the living knowledge and skills passed down through generations. Daily practice plays a vital role in revitalizing ICH by fostering continuous learning and improvement. However, limited resources and accessibility pose significant challenges to sustaining such practice. Virtual reality (VR) has shown promise in supporting extensive skill training. Unlike technical skill training, ICH daily practice prioritizes cultivating a deeper understanding of cultural meanings and values. This study explores VR's potential in facilitating ICH daily practice through a case study of Traditional Chinese Flower Arrangement (TCFA). By investigating TCFA learners' challenges and expectations, we designed and evaluated FloraJing, a VR system enriched with cultural elements to support sustained TCFA practice. Findings reveal that FloraJing promotes progressive reflection, and continuous enhances technical improvement and cultural understanding. We further propose design implications for VR applications aimed at fostering ICH daily practice in both knowledge and skills.

\end{abstract}


\begin{CCSXML}
<ccs2012>
   <concept>
       <concept_id>10003120.10003121.10011748</concept_id>
       <concept_desc>Human-centered computing~Empirical studies in HCI</concept_desc>
       <concept_significance>500</concept_significance>
       </concept>
 </ccs2012>
\end{CCSXML}

\ccsdesc[500]{Human-centered computing~Empirical studies in HCI}

\keywords{Virtual Reality, Intangible Cultural Heritage, Daily Practice, Traditional Chinese Flower Arrangement}
\begin{teaserfigure}
  \centering
  \includegraphics[width=0.9\linewidth]{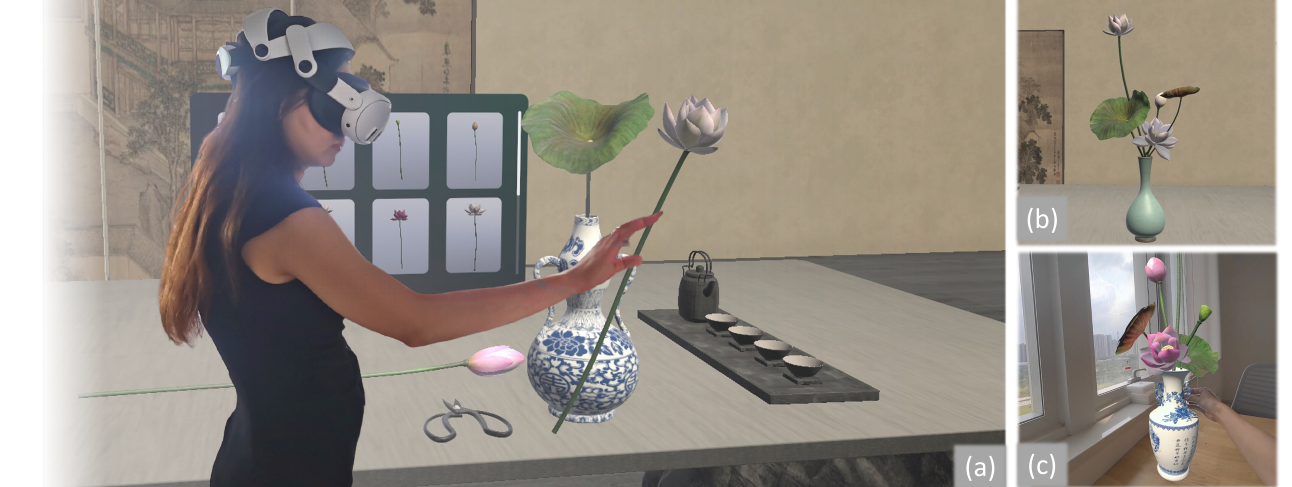}
  \caption{
  Traditional Chinese flower arrangement (TCFA) is one intangible cultural heritage that requires daily practice to refine skills and learn related culture. We designed FloraJing, a VR application that addresses the challenges in TCFA daily practice by providing rich cultural environments and accessible materials and tools. \textbf{(a)} In FloraJing, a user arranges flowers with natural hand gestures in a Chinese tea room. \textbf{(b)} The user saves the TCFA work and captures it. \textbf{(c)} In mixed reality mode, the user interacts with the digital TCFA within the physical environment.}
  \Description{This figure contains three parts, (a), (b) and (c). Figure (a) shows that in FloraJing VR, a user engages in selecting and arranging lotus flowers and leaves with natural hand gestures in a Chinese tea room, practicing TCFA skills. There are virtual TCFA materials in the virtual tea room. In Figure (b), there is finished TCFA work in the virtual tea room. Figure (c) shows that a user places the virtual TCFA work in their real-world surroundings.}
  \label{fig:teaser}
\end{teaserfigure}



\maketitle

\section{INTRODUCTION}

\rv{Intangible cultural heritage (ICH) includes the traditions and living expressions passed down from our ancestors to future generations, represented in five domains, such as traditional craftsmanship, performing arts, and knowledge and practices concerning nature and the universe \cite{WhatisIntangible}. ICH serves as a bridge connecting past, present, and future, playing a vital role in the preservation of cultural diversity amid growing globalization \cite{livingintangible_2011, importance_2016}. Its significance lies not in the cultural forms themselves, but in the living knowledge and skills transmitted through these forms across generations \cite{definition_2010, WhatisIntangible, digitizing_2022}.} ICH knowledge and skills are inseparable: skills objectify knowledge, while knowledge provides skills with their value. ICH knowledge is acquired and understood through long-term, accumulated experience gained via embodied, hands-on skill practice \cite{digitizing_2022, deliberate, Apprenticessun_research_2021}.

However, traditional \textit{\rv{ICH knowledge}} and \textit{skill transmission} methods, such as apprenticeships, face challenges due to being labor-intensive, time-consuming over extended periods, and constrained by geographical limitations \cite{zhou2024integrating}. These factors often lead younger generations to lose interest or discontinue their learning \cite{unesco}. 
While local residents can participate in ICH workshops, challenges such as limited financial support and irregular scheduling hinder sustained practice \cite{locationsmuka2016intangible}. ICH practices were once an integral part of daily life routines \cite{kurin2007safeguarding}, and reviving them through daily practices in contemporary life presents a promising prospect. However, societal changes, including the limited availability of local resources and rising production costs \cite{unesco}, have made sustaining ICH daily practice increasingly difficult.
While prior HCI research has explored various digital technology solutions (e.g., motion capture, interactive digital storytelling, and integrating digital tools with traditional crafts production) for ICH preservation \cite{danceanalysis, Storytellingspace, HybridCraft}, much of this work has focused on singular experiences rather than fostering continuous daily practice to revitalize ICH. This highlights the need to explore the challenges ICH learners face in sustaining their daily practice and develop innovative solutions that seamlessly integrate ICH practice into daily life.

\rv{Previous studies on virtual reality (VR) have demonstrated its potential to support continuous and extensive skill practice by providing virtual environments and interactions that closely simulate real-world conditions. Its effectiveness has been proven in fields where real-world constraints limit regular practice, such as surgical skills \cite{knottyingskills, bonetechnical}, motor skills \cite{motorskill1, cricket}, and personal safety skills \cite{ccakirouglu2019development}. In these areas, trainees are expected to perform behaviors that closely align with the demonstrated or desired standards. However, compared to technical skill training, ICH practice places greater emphasis on fostering interest, understanding, and creativity within the concerned communities \cite{bakar2011intangible}. Achieving this goal requires reflecting on concrete ICH practices, using methods such as active imagination to abstract cultural meanings from specific technical rules \cite{activeimaginationunderstanding}, and continuing to actively participate in ICH practices. Given the distinctive nature of ICH, it is crucial to explore how to design VR experiences that effectively facilitate ICH daily practice.

In this study, we take Traditional Chinese Flower Arrangement (TCFA) as a case study, aiming to explore design strategies that leverage VR's potential to support and enhance ICH daily practice.} TCFA is a craft-based ICH deeply rooted in Chinese philosophy and literature \cite{xu_inheritance_2023}. \rv{These cultural influences are reflected in its styles, principles, and techniques \cite{caiArtChineseFlower2017d, lianying_art_2020},} such as the Confucian idea of \textit{benevolence},  which is expressed in TCFA through the practice of cherishing flowers and decorating them minimally \cite{xu_inheritance_2023}. Another key example is TCFA's pursuit of yijing, an essential aesthetic in China's poetic traditions, which blends emotions and scenery to create a lasting, poetic appeal, often through techniques like liubai (intentional blank spaces)  \cite{wang2017aesthetics, ideological}. TCFA was once one of \textit{the Four Arts of Living} practiced by Chinese for physical and mental well-being \cite{li2002chinese}, \rv{but declined since the Qing Dynasty (1636–1912) due to invasions and economic recessions \cite{caiArtChineseFlower2017d}, nearly disappearing from folk traditions \cite{lianying_art_2020}. Today,} TCFA is experiencing a resurgence, especially among younger generations drawn to traditional culture and an aesthetically enriched lifestyle \cite{ZhangJie}. As a form of plastic art, TCFA demands continuous practice for learners to master materials, tools, aesthetic principles, and underlying cultural knowledge and philosophical reflections \cite{caiArtChineseFlower2017d}. 
However, the high cost and limited availability of essential materials (e.g., woody plants and Chinese-style vases) hinder its daily practice \cite{ZhangJie}.

\begin{figure*}[t]
    \centering
    % \includesvg[width=0.95\linewidth]{Figures/flow chart.svg}
    \includegraphics[width=0.9\linewidth]{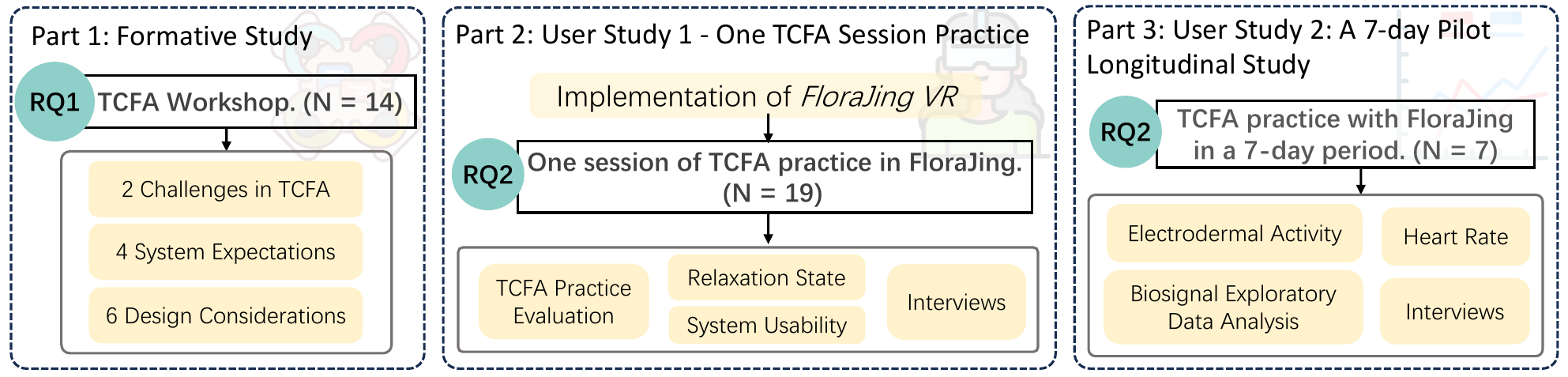}
    \caption{\rv{Part 1 investigates the current challenges in TCFA learners' daily practice and derives design considerations; Part 2 focuses on evaluating whether the VR system can support learners' daily practice; Part 3 is a 7-day pilot longitude study for assessing whether and how FloraJing could sustain daily practice over time.}}
    \label{fig: research-flow}  
    \Description{.}
\end{figure*}

\rv{Our research flow is shown in Fig. \ref{fig: research-flow}.} In this study, to address \textbf{RQ1}: What specific challenges do TCFA learners face in their daily practice, and what technological solutions do they envision?, we conducted three workshops, which included a TCFA practice session and a focus group discussion. We open-coded the results and identified key barriers to daily TCFA practice, such as the lack of a cultural atmosphere in physical environments and the \rv{inability to revisit and reflect on practice outcomes}. From these findings, we derived \rv{six} design considerations (DCs). Based on these DCs, we designed and implemented FloraJing, a hand-interactive VR system aimed at supporting TCFA learners' daily practice.
To address \textbf{RQ2}: How effectively do our VR designs support and sustain TCFA daily practice, and what improvements can be made in the future?, we conducted an evaluation study with learners of varying levels of TCFA expertise. This included a TCFA practice session to assess whether FloraJing could support learners' daily practice. \rv{Following this, we carried out a pilot 7-day longitudinal study to evaluate its effectiveness in sustaining TCFA daily practice over time.} Key findings indicate that FloraJing supports and sustains TCFA daily practice by \rv{(1) facilitating progressive reflection and improvement based on previous TCFA work records, (2) enabling virtual creative experiences that can be transferred to real-life TCFA practice, and (3) enhancing understanding of TCFA within an immersive, culturally enriched environment.}

We make two primary contributions:
(1) We design and develop a VR application to support TCFA learners in their daily practice, \rv{both in cultural knowledge and skill, offering a novel approach to help preserve ICH,}
(2) We propose design implications for future researchers to develop effective VR applications to support ICH daily practice.

\section{RELATED WORK}
\subsection{ICH Preservation with Digital Technologies in the HCI Community}

A growing body of work in the HCI community has focused on digital innovations aimed at preserving ICH. The digital preservation lifecycle of ICH resources can broadly be divided into three key stages: recording, representing, and reviving \cite{digitizing_2022}. In the recording phase, techniques similar to those used in tangible heritage (TH) preservation, such as photography, 3D scanning, and modeling, are employed to archive ICH objects and artifacts \cite{ShanghaiStyleLacquerware}. However, unlike TH, which is centered around physical objects, the cultural expressions of ICH are defined through tacit reliances and embodied practices \cite{digitizing_2022}. Therefore, motion capture systems and computer vision methods are employed for movement acquisition, recognition, and reproduction within ICH, such as the sequence trajectories of traditional folk dances \cite{danceanalysis} or the manufacturing process in traditional craftsmanship \cite{traditionalcrafts}. 

% The recorded data is often visualized in VR or augmented reality (AR).

For ICH representing, HCI researchers primarily use VR-based \cite{huaer, Accessibility, touch} or AR-based \cite{book, OperARtistry} systems to create authentic and immersive experiences. These immersive systems often integrate interactive digital storytelling \cite{huaer, yujian, Storytellingspace} and game-based learning \cite{role-play} approaches to raise awareness and enhance public engagement with ICH. For example, Liu et al. developed an interactive VR system to engage audiences with the traditional oral performance of Hua'er, which improved users' understanding of its cultural significance through participatory performance and knowledge-based gameplay \cite{huaer}. These interactive systems have also been shown to strengthen family ties \cite{FamilyWell-being} and foster intercultural exchange \cite{interculturalheritage}.   These works primarily focus on providing short-term, single experiences rather than long-term integration into daily life.

% Similarly, AR-based systems, such as those used for Chinese Opera makeup, enhance user engagement and learning by offering step-by-step guides for beginners \cite{OperARtistry}.

 Reviving ICH through technology primarily focuses on creatively integrating digital tools with traditional crafts \cite{Hunter-Gatherer, bamboo, hairymonkey}. For instance, Jacobs and Zoran worked with a hunter-gatherer community, combining digital tools with ostrich eggshell jewelry crafting to incorporate modern design into ancient practices, enriching communal life and sustaining cultural practices \cite{Hunter-Gatherer}. Research on supporting practice within ICH has been less explored but is essential for reviving ICH in modern contexts. For example, Yu et al. developed a mixed reality learning environment for Guqin, a traditional Chinese musical instrument, which significantly lowered the entry barriers for learners and enhanced their practice performance \cite{guqin}. This study illustrates how technology can empower ICH practices and aid in the preservation of ICH traditions at risk of fading away.

 Much of the digital preservation research on ICH in the HCI community has focused on the recording and representing stages, with fewer studies addressing the challenges of reviving ICH. Reviving requires the continuous adaptation and creative practice of ICH in contemporary contexts \cite{HybridCraft}. Daily practice, a key aspect of reviving ICH, integrates ICH into the everyday lives of practitioners, keeping these traditions alive and dynamic. However, factors like diminishing apprenticeships, and globalization have negatively impacted ICH daily practices \cite{unesco}, making it crucial to explore how digital technologies can support the daily practice of ICH in modern life.

\subsection{\rv{VR for Supporting Skill Practice}}

\rv{VR has increasingly emerged as a high-fidelity simulation modality for skill practice and acquisition, particularly in fields where real-world resource constraints hinder regular training. Its effectiveness in supporting skill development and maintenance has been demonstrated across diverse fields requiring continuous, intensive training, such as surgical skills \cite{knottyingskills, bonetechnical}, motor skills \cite{motorskill1, cricket}, and mindfulness practice \cite{zenvr_2022, mindfulnesspractice}.\par

Virtual reality offers several technical advantages for skill practice and development. First, 3D visual representation in VR offer richer information that closely approximates real-world environments, enhancing skill practice by allowing users to view scenarios from multiple angles \cite{simulationmodality}. For example, in surgical training, computer-generated 3D models in VR enabled medical students to better locate abnormalities and plan surgeries with greater precision compared to 2D methods \cite{knottyingskills, comparison2-dimensional}. Additionally, compared to other skill practice methods, VR provides a learner-centered, embodied experience that promotes independent practice. Feinberg et al. developed a learner-centered VR application to support progressive meditation practice, showing significant improvements in both mindfulness and self-reported meditation abilities \cite{zenvr_2022}. The embodied affordances of VR, such as gesture-based interactions within a 3D space, have been found to positively influence real-world learning outcomes by reinforcing physical actions performed in virtual environments \cite{Embodied}. Moreover, VR with multisensory integration, such as haptic feedback, can guide users to avoid unnecessary actions and ultimately reduce their cognitive load, as demonstrated in the training of dental anesthesia technical skills \cite{collacco2021immersion}.

However, most VR applications for skill training largely prioritize technical skill development, requiring trainees to replicate behaviors that align with predefined standards. Outcomes in such contexts are typically quantified through objective metrics like accuracy or speed. In contrast, ICH practice emphasizes deepening understanding, and cultivating interest and creativity within the communities \cite{bakar2011intangible}. While VR has been found to enhance enthusiasm, creativity, and emotional engagement in art practice \cite{som2023virtual, ArtPractice}, further exploration is needed to understand how VR can be leveraged not only for training the practical skills of ICH but also for fostering the rich cultural and communal knowledge embedded within these practices. In this study, we take TCFA as a case study to investigate strategies for maximizing VR's potential in supporting ICH daily practices.}

\subsection{\rv{VR for Cultural Heritage Preservation}}

\rv{In recent years, VR has emerged as a valuable tool for both cultural heritage (CH) and ICH preservation by improving accessibility and offering realistic, immersive, and interactive experiences. Various immersive forms, such as Cave Automatic Virtual Environment (CAVE) \cite{cavechristou2006versatile}, 360° videos \cite{videostory2018vr, touch}, VR headsets \cite{li2023diantea, 2018tales}, or their combinations, have been employed for this purpose.

VR's immersive capabilities have been utilized to create holistically immersive environments for showcasing digitized ICH works, such as 3D animations derived from recorded dance movements \cite{traditionalcrafts, 2020creative}. Additionally, the embodied affordances of VR enable participants to take on roles within CH scenarios, allowing them to engage in storytelling and practices from a first-person perspective. This interactive approach sparks interest in traditional events and encourages users to transition from passive observers to active learners \cite{2018tales, li2023diantea}. Non-playable characters (NPCs) in VR can serve as realistic agents for guidance and knowledge dissemination, significantly boosting users' engagement and interest in CH \cite{huaer, vosinakis2018dissemination}. Moreover, integrating VR with multimodal feedback significantly heightens the sense of presence while enriching the experience with additional CH information \cite{touch}. For example, Sun et al. incorporated olfactory and wind feedback to convey invisible cues. These multisensory elements also served as task annotations, provided spatial hints, and helped users gain a deeper understanding of the virtual CH environment \cite{restoringsun2024}. These findings demonstrate that VR can promote active participation and cultural exploration in CH preservation. These insights inform our system design to support TCFA learners in their daily practice effectively using VR.}

\section{FORMATIVE STUDY}
We conducted three workshops with 14 participants of varying TCFA expertise to identify specific challenges and expectations related to daily TCFA practice \textbf{(RQ1)}, which inform our design of a supportive system. All user studies followed the standard ethical procedures of the research institute. 

\subsection{Workshop} \label{workshop}

To better understand TCFA before conducting the workshops, two authors read TCFA related books \cite{lianying_art_2020, li2002chinese} and attended courses \rv{to gain} a foundational understanding of TCFA. Every workshop included a TCFA practice session and a focus group discussion. The practice session was designed to help participants recall more detailed aspects of their daily TCFA practice and to foster connections among participants. The focus group discussion aimed to understand the detailed challenges. In total, we held three workshops, each lasting between 1 to 1.5 hours. Participants received a \$15 compensation for their participation.

We received support from three TCFA instructors throughout the workshops. They assisted in participant recruitment by sharing the recruitment poster, provided guidance and support on venue setup and material preparation for TCFA practice sessions, and also joined the workshops as participants, offering valuable insights during the focus group discussions.

\subsubsection{Participants}
We shared a recruitment poster on social media platforms and TCFA online communities to recruit participants with diverse daily practice habits and varying levels of expertise in TCFA. The poster included a questionnaire to gather details on participants' daily practice routines and self-reported learning levels (classified into five levels: Novice, Beginner, Competent,
Proficient, and Expert). In total, we recruited 14 participants (see Table~\ref{table:participant-workshop}).
% \rv{with each workshop including one TCFA instructor as an expert and three to four learners of varying expertise levels} 

\begin{table*}[t]
    \centering
    \caption{Information of Participants in TCFA Workshop. The table provides participants' demographics and their relevant experience in TCFA. \textit{Years of Experience} refers to the duration from when a participant began learning TCFA until the experiment. \textit{Total Time Spent} refers to cumulative amount of time dedicated to practicing TCFA. $*$ notes that P1, P5 and P10 are TCFA instructors who helped us during the workshop and joined the workshop as participants.}
    \label{table:participant-workshop}
    \Description{This table shows demographics and relevant experience of participants in our user study. ID, age, sex, prior experience with Traditional Chinese flower Arrangement (TCFA) and frequency of daily practice are included in this table.}
    \resizebox{0.7\textwidth}{!}{
    \begin{tabular}{cccllll}
   \hline
 &  &  & \multicolumn{3}{c}{\textbf{Prior TCFA Experience}} &  \\ \cline{4-6}
\multirow{-2}{*}{\textbf{ID}} & \multirow{-2}{*}{\textbf{Age}} & \multirow{-2}{*}{\textbf{Sex}} & Skill Level & Years of Experience & Total Time Spent & \multirow{-2}{*}{\textbf{\begin{tabular}[c]{@{}l@{}} Practice Frequency \end{tabular}}} \\ \hline
\multicolumn{7}{l}{\textit{Group 1}} \\ \hline
P1$*$ & 29 & F & Expert & 7 years & \textgreater 200 hours & Every week \\
P2 & 26 & F & Beginner & 3 months & 10 hours & Every month \\
P3 & 27 & M & Beginner & 1 month & \textless 5 hours & Every month \\
P4 & 22 & F & Competent & 2 year & 40 hours & Every 2 months \\  
\hline
\multicolumn{7}{l}{\textit{Group 2}} \\ \hline
P5$*$ & 35 & F & Expert & 7 years & \textgreater 200 hours & Every week \\
P6 & 24 & F & Beginner & 2 months & 5 hours & Every month \\
P7 & 26 & F & Competent & 4 months & 60 hours & Every week \\
P8 & 28 & M & Competent & 7 months & 50 hours & Every 2 weeks \\
P9 & 20 & M & Proficient & 1 years & 120 hours & Every week \\ \hline
\multicolumn{7}{l}{\textit{Group 3}} \\ \hline
P10$*$ & 37 & M & Expert & 5 years & \textgreater 200 hours & Every week \\
P11 & 33 & F & Beginner & 1 year & 15 hours & Every month \\
P12 & 27 & F & Beginner & 2 years & 15 hours & Every two months \\
P13 & 28 & M & Competent & 1 year & 70 hours & Every two weeks \\
P14 & 32 & F & Proficient & 5 years & 100 hours & Every month \\ \hline
\end{tabular}}
\end{table*}

\subsubsection{Procedure}\leavevmode

1) \textit{Introduction.}
We first obtained informed consent from each participant, including permission for audio recordings and photos during the workshop. Next, we provided a brief introduction explaining the study's objectives and workshop procedures. Two researchers facilitated the session—one managed coordination, while the other documented the process.

% The paper lacks clarity on crucial aspects of the research, such as a complete list of materials—both those used in the workshop and those developed in the prototype by the authors. Some materials are mentioned, but it's unclear if this is an exhaustive list or just a selection of examples. Additionally, there is no explanation of how the materials were chosen or what criteria guided their selection.

\begin{figure}[t]
  \centering
  \includegraphics[width=0.7\linewidth]{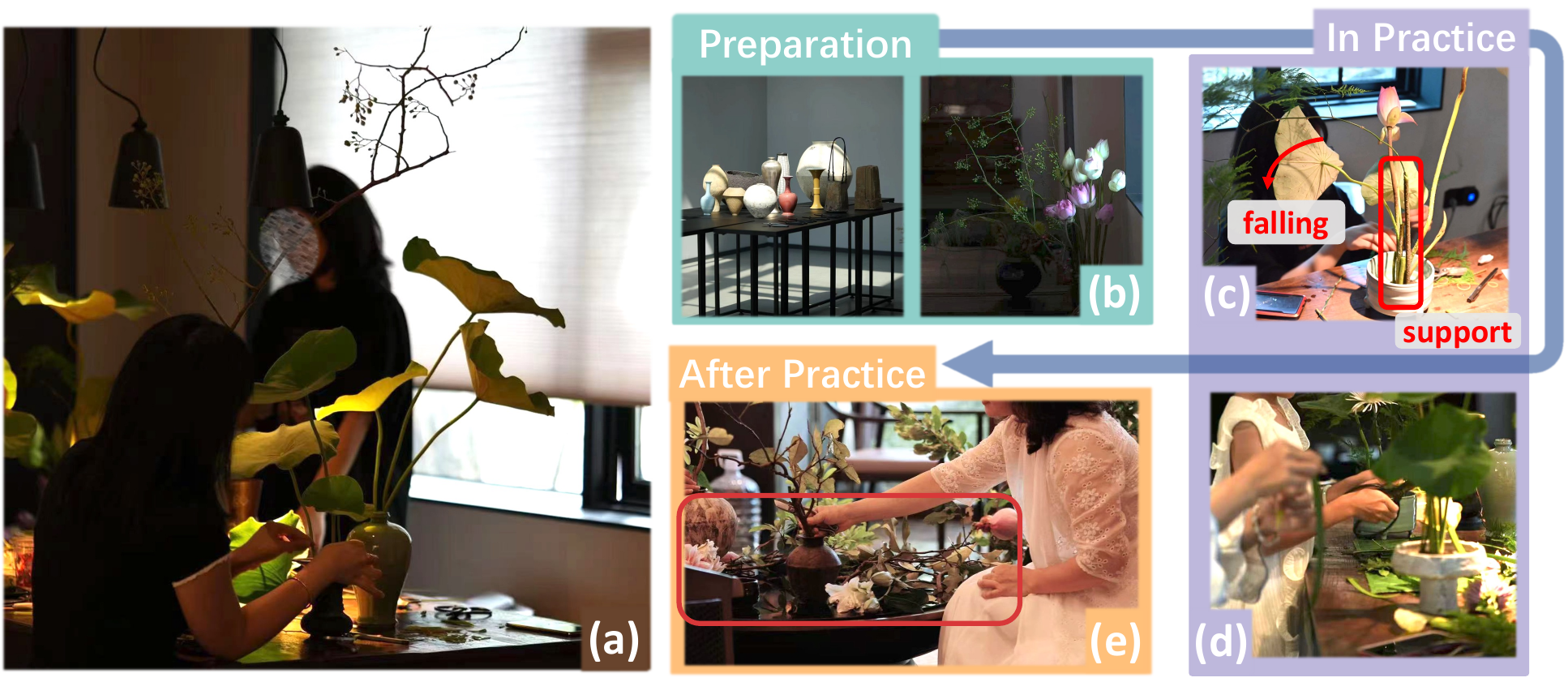}
  \caption{
  \textbf{(a)} One of the TCFA workshop scenes in our study. 
  \textbf{(b)} Difficulties in preparing the TCFA's material. 
  % All participants expressed that preparing materials, such as Chinese-style vessels and woody flowers, is the most challenging part of the daily practice of TCFA. 
  \textbf{(c)} Frustration related to shape stability. 
  % Due to gravity or wilting, maintaining the stability of arranged flowers and leaves during the process is very challenging. Wire or poles are needed to support the plants. 
  \textbf{(d)} Anxiety of irreversible pruning decisions. 
  % The irreversible nature of pruning creates significant psychological pressure in daily practice. 
  \textbf{(e)} Burden of post-practice cleanup. 
  % After the practice, significant time and effort are required in clearing the trimmed materials and the site.
  }
  \label{fig:workshop-challenge}
  \Description{Figure(a) shows the scene of workshop in our study. It is a tea-room-like studio for flower arrangement. In figure(b), it shows the Chinese-style vessels and woody flowers needed for the TCFA. Figure(c) shows a learner sitting near the desk, using wire and poles to support the flowers and the leaves from failing. Figure(d) shows learners use scissors to prune the plants, creating lots of irreversible materials. In the last picture, figure(e) shows a messy desk with many trimmed materials, and a learner is cleaning the desk.)}
\end{figure}

2) \textit{TCFA Pratice Session.} Participants took part in a free-form TCFA practice session. To ensure convenience, we consulted TCFA instructors and participants to identify essential materials for daily practice and reserved a spacious tea room based on their recommendations (shown in Fig.\ref{fig:workshop-challenge}(a)). We provided tools like pruning scissors, floral pin holders, traditional Chinese vases, and seasonal classic materials, including lotus flowers, and leaves (see Fig.~\ref{fig:workshop-challenge}(b)). This session aimed to help participants immerse themselves in the practice of TCFA, encouraging them to reflect on their daily routines and experiences.

3) \textit{Focus Group Discussion.} 
After practice, the focus group discussion took place beside the flower arrangement table, enabling participants to interact with the tools and refer to their work to better express themselves. Since the \textit{TCFA Practice Session} has functioned as an icebreaker, the focus group discussion was divided into two main parts. The first part (25 minutes) focused on challenges in daily TCFA practice. Participants were asked questions such as: \textit{What challenges do you face when practicing TCFA in your daily life?} and \textit{How do these challenges affect your daily practice?} The second part focused on a discussion of desired technological solutions (20 minutes). Questions included \textit{What features or tools would you like to see in a technological solution that supports your TCFA daily practice?} and \textit{What aspects of the TCFA experience are most important to you that should be preserved or enhanced?}

% , the complete 访谈大纲 is presented in Appendix \ref{XX}.

\subsection{Data Analysis}
The data analysis aims to identify challenges and expectations of current TCFA practice, and inform the key features that better support this process. Our data consisted of observational note and audio recordings from the workshop. 
The audio recordings were transcribed into a text script from the workshop. 
To analyze the data, we followed the thematic analysis \cite{oktay2012grounded} method. 
Initially, four researchers thoroughly reviewed the script to gain a comprehensive understanding of the content. Subsequently, two authors conducted the open-coding process independently after familiarizing themselves with the data. 
During weekly meetings, all members deliberated on the interpretations of the data, and reached a consensus on the final coding results.

\subsection{Findings} \label{workshop findings}
% Compared to the teacher daily self-practice frequency (both about 1/3 days), the student is low （max： 1/2 week, min:1/7 month. 
% Many students have
% expressed that 日常中式插花训练耗时长，买花材花器昂贵，中式插花的花材例如木本花材难买等阻碍了他们中式插花的日常训练。
% 1. 练习的成果无法再次修改反思

% 2. 插花过程前后工作的繁杂

% 材料 清洁

% 3. 日常插花环境文化氛围的缺乏

% 4. 插花过程中漫长的固定工序导致时间安排的困难

% The frustration related to shape stability.
% The anxiety of irreversible pruning decisions. 
\subsubsection{Challenges in the \rv{TCFA} Daily Practice} \leavevmode 
 \par\textbf{\rv{Physical} Environmental Constraints on Cultural Immersion.}  All participants recognized that cultural immersion through environmental settings enhances their TCFA practice experience, helping them materialize abstract aesthetic concepts and philosophical reflections, such as yijing. However, participants (N=7) noted that everyday physical environments, such as cramped or noisy spaces, often fail to provide sufficient cultural immersion, affecting their mood and diminishing interest in daily practice. As P4 mentioned, “I currently live in a rented place with a small and cluttered space, so I don’t feel the same sense of being immersed and relaxed as I do in class, which makes me less inclined to practice at home.” Regarding TCFA practice settings, three instructors recommended selecting venues that align with Chinese aesthetics (e.g., temples, Zen-inspired tea rooms), as they typically choose such spaces for their classes to better convey the sense of Chinese beauty to students. \textbf{- Design Consideration (DC)1: The prototype should incorporate elements that create a culturally immersive environment, even in physically constrained spaces.}

\rv{\textbf{Peripheral Tasks External to the Creative Process.}
Participants reported that tasks outside the core flower arrangement session, such as acquiring materials (N=14) and post-practice cleanup (N=8), posed significant challenges to maintaining regular daily practice (shown in Fig.~\ref{fig:workshop-challenge}(b)\&(e)). These tasks diminished enthusiasm, with some participants opting for classes as a complete substitute for daily practice (N=5).} \rv{For material acquisition,} TCFA often requires ligneous plants with distinctive lines to convey energy and rhythm, which are scarce in flower markets. As P8 explained, \textit{``Sometimes I stop by after work to buy flowers, but I mostly see flowers like roses. It's rare to come across traditional TCFA flowers like lotus or asparagus fern."} 
 \rv{Regarding cleanup,} participants reported spending between 20 minutes to over an hour on the task, significantly extending the overall activity duration. This was particularly challenging for those with busy schedules. \rv{\textbf{- DC2: The prototype should reduce the burden of peripheral tasks to maintain learners' enthusiasm and accessibility for TCFA daily practice.}}

\subsubsection{System Expectations}\leavevmode
\par\rv{\textbf{Enabling Progressive Revisions of TCFA Works.} All participants expressed regret that their previous TCFA works could not be revisited or refined when new inspiration or learning insights emerged. They hoped for a system that could preserve their arrangements in a 3D form, allowing them to freely view and modify their works. In real-life practice, flower arrangements inevitably wilt and fade, leaving participants with only photos to document their creations. This reliance on static images makes it difficult to refine past works when new ideas arise, hindering continuous reflection and improvement. As P13 mentioned, \textit{``Tracking practice progress is very important for my learning. When I work on my next TCFA arrangement, I often review my past creations. When I have new ideas, I really wish I could revise them, but it's just not possible."} \textbf{- DC3: The system should enable users to save their TCFA works in a modifiable 3D format, supporting continuous reflection, improvement, and creative exploration.}}

\textbf{Embodied Interaction for Authenticity.} All participants expressed a desire for the prototype system to involve physical behaviors, mainly hand gestures, to conduct \rv{TCFA} in a more intuitive and natural way. \rv{As P1 (TCFA instructor) shared her experience in learning \rv{TCFA},\textit{``At first, I felt clumsy and could not create what I envisioned. But through continuous practice, I reached a state where my hands moved in harmony with my mind, achieving a seamless integration of body and spirit.''}} \textbf{- DC4: The prototype should incorporate natural and intuitive interaction methods that closely mimic the real-life experience of \rv{TCFA}.}

\rv{\textbf{Connect the Virtual Work with Reality.}}
Most participants (N=9) expressed a desire for their TCFA creations to have a connection with their real lives, fostering a stronger emotional bond. As P6 noted, \textit{``When I played a flower arrangement game, I created a beautiful piece, but it could only stay on the computer. I wished I could see it on my table at home—then it would feel truly special, and my effort would have more meaning.''} \rv{\textbf{- DC5: The system should bridge the virtual and physical worlds to connect the virtual TCFA works with reality.}}

\rv{\textbf{Transcending Real-World Physical Limitations.}} \rv{Participants (N=9) expressed a desire for the system to offer features that go beyond real-world limitations to enhance their creative processes. Two primary challenges were identified as obstacles to their creative expression.} \rv{First, the instability of long-stemmed plants commonly used in TCFA made it particularly difficult to maintain the desired shape, often necessitating the use of wires or poles for support. (shown in Fig.\ref{fig:workshop-challenge}(c)), which led to learners' frustration and reduced interest. Second, }the irreversible nature of pruning created a fear of making mistakes, which could potentially ruin the arrangement or waste flowers (shown in Fig.\ref{fig:workshop-challenge}(d)). \rv{\textbf{- DC6: The system should provide features that transcend real-world physical constraints to support unrestricted creative exploration and reduce frustration.}}

\section{PROTOTYPE DESIGN AND IMPLEMENT}
Based on the formative study findings and the derived DCs, we proposed FloraJing, a VR-based prototype designed to support daily practice for learners of TCFA. \rv{The name \textit{FloraJing} was chosen to reflect the pursuit of yijing in TCFA, emphasizing the deep aesthetic and cultural experience that the system aims to facilitate.} In this section, we detail the design and implementation of FloraJing, including its core features, user interface, interaction mechanisms, and how these elements align with the needs and preferences of TCFA learners identified in our formative study.

\begin{figure*}[t]
  \centering
  \includegraphics[width=0.9\linewidth]{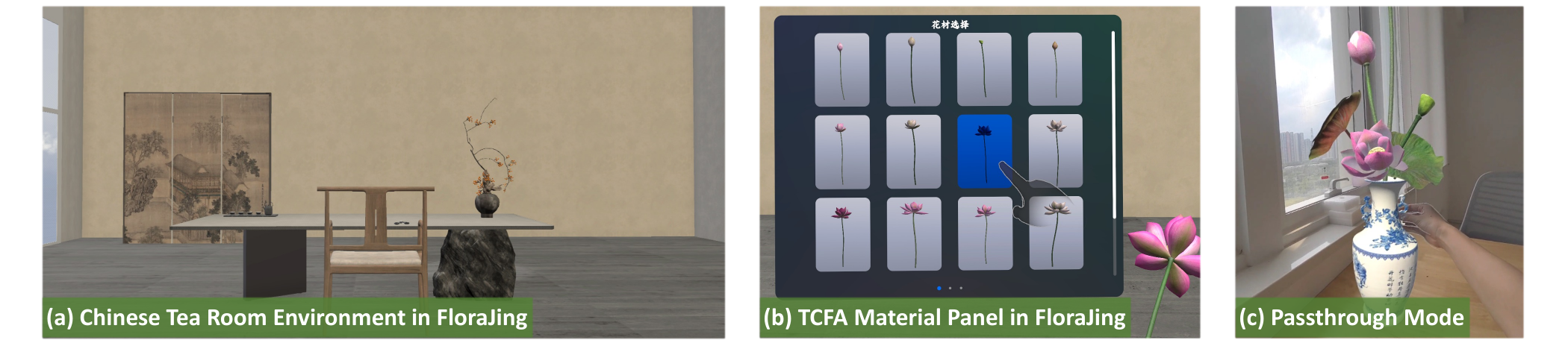}
  \caption{
  \textbf{(a)} A virtual Chinese tea room served as the setting to enhance cultural immersion. 
  \textbf{(b)} Material selection panels in FloraJing, including \textit{Flower Panel}, \textit{Leaf Panel} and \textit{Vessel Panel}. 
  \textbf{(c)} Mixed reality passthrough mode in FloraJing.}
  \label{fig:scene-panel-pass}
  \Description{
  Figure-a shows a representative TCFA environment, specifically a tea room, was utilized as the VR scene in FloraJing. 
  Figure-b shows the three material selection floating Panels in FloraJing, including \textit{Flower Panel - 14 lotus models}, \textit{Leaf Panel - 16 lotus leaf models} and \textit{Vessel Panel - 12 Chinese vessel models }. 
  Figure-c shows the passthrough mode in FloraJing. Users can enjoy their creations in the real world while in mixed reality passthrough mode.
  }
\end{figure*}

\subsection{Overview}
We developed FloraJing with Unity3D (2022.3.15f1-LTS)\footnote{Unity: https://unity.com/releases/2022-lts} and implemented it with an Oculus Quest 3\footnote{Meta Quest 3: https://www.meta.com/quest/quest-3/} VR headset on a desktop computer with AMD 3500X, 16GB RAM and RTX3090 GPU. Meta XR All-in-One SDK\footnote{Meta XR All-in-One SDK: https://assetstore.unity.com/packages/tools/integration/meta-xr-all-in-one-sdk-269657} ver.66 was included to provided VR features.

The selection of a theme for FloraJing was guided by the principles and traditions of TCFA. TCFA emphasizes harmony with nature, and the reflection of seasonal changes, often referred to as ``timely flower arrangement'' \cite{lianying_art_2020}. In ancient China, arranging lotus flowers was a traditional summer practice, considered a way to stay cool during the hot months. The lotus, in Chinese symbolism, represents purity and noble character, as it is said to ``emerge unstained from the mud''. Since our experiment was conducted in the summer, we chose the lotus as the practice theme.

\subsection{System Design and Implement}

\textbf{Virtual \rv{Tea Room Design}.} According to \textbf{DC1}, we utilized a virtual tea room environment with traditional Chinese-style decoration to create a culturally immersive TCFA environment (shown in Fig. \ref{fig:scene-panel-pass}(a)).

\begin{figure*}[t]
  \centering
  \includegraphics[width=\linewidth]{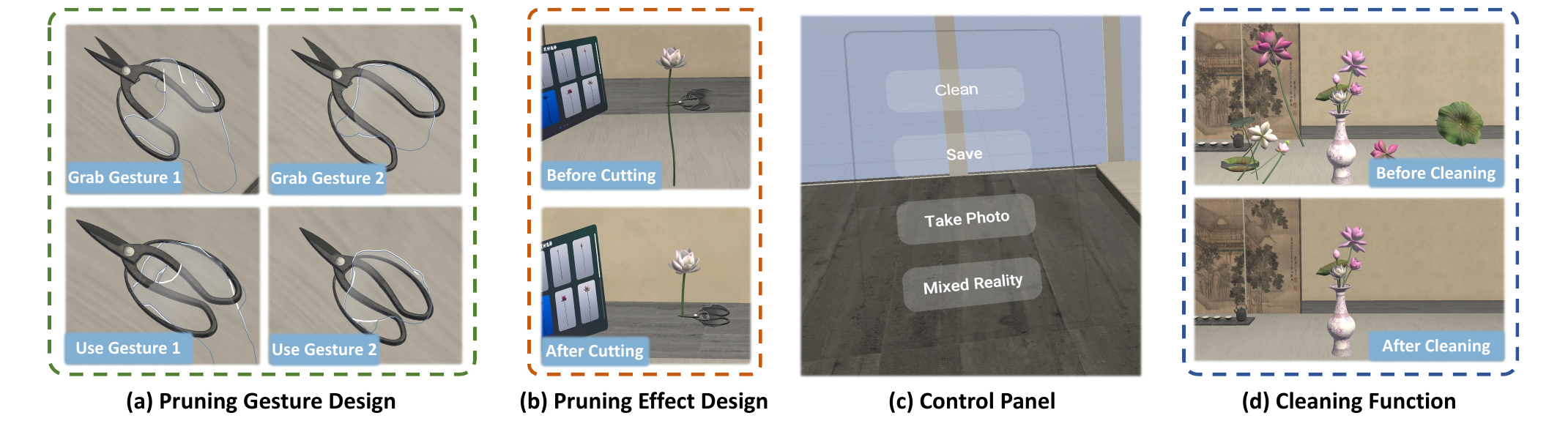}
  \caption{\textbf{(a)} Pruning gesture design. 
  % We recorded two hand gestures to accommodate different user usage habits.
  \textbf{(b)} Pruning effect design, with trimmed parts disappearing automatically. 
  \textbf{(c)} FloraJing control panel. 
  \textbf{(d)} The cleaning function removes unwanted materials.}
  \label{fig:hand-pruning-save-clean}
  \Description{Figure-a shows the Pruning gesture design, featuring two grab gestures and two usage gestures. We recorded two hand gestures to accommodate different user usage habits.
  Figure-b shows Pruning effect design. users grasp the scissors and utilize them to prune the materials. The trimmed part will automatically disappear. 
  Figure-c shows the control panel in FloraJing, offering users functionalities to clean the scene, save the works, take photos of the works and activate the mixed reality mode. 
  Figure-d shows the cleaning function in FloraJing. After clicking the clean button, the unwanted materials that are not in contact with the vessel will be removed.
  }
\end{figure*}

\textbf{Gesture Interaction and Pruning Features.} To achieve natural and intuitive interaction (\textbf{DC4}), we applied hand gesture interaction in FloraJing and attached VR grabbable modules to all TCFA materials, allowing users to perform natural gestures to freely grab, rotate and release the materials as they would in the real world. For the Pruning Feature, we used the \textit{hand gesture recorder} in Unity to capture different hand gestures associated with scissor use, enabling a realistic simulation of the pruning process. Fig. \ref{fig:hand-pruning-save-clean}(a) showcases the recording of two hand gestures for grabbing and using scissors, designed to adapt to different users' usage patterns.

\rv{\textbf{Material Selection and Cleaning Features.} According to \textbf{DC2}, we used categorized lists to provide users with easy access to TCFA materials and provided a variety of models in different styles, colors, and shapes to give participants more creative freedom, including 12 different virtual Chinese vessels, 14 lotus flowers, and 16 lotus leaves, as shown in Fig. \ref{fig:scene-panel-pass}(b). We also designed the unwanted parts disappear after being trimmed, as shown in Fig. \ref{fig:hand-pruning-save-clean}(b). Additionally, when users press the clean button on the control panel in Fig. \ref{fig:hand-pruning-save-clean}(c), plant models that are in contact with the chosen vessel's collider will remain, while others will be removed, as shown in Fig. \ref{fig:hand-pruning-save-clean}(d).} 

% \textit{Considerations and Trade-offs of the Gravity System.}
% % \textit{Cleaning Features.}
% \subsubsection{Cleaning Features}
%save bottom, 保留collider，其余清除
\textbf{Artwork Saving, Photo Capture\rv{, and Mixed Reality} Features.} In line with \textbf{DC3}, we designed the save, and photo capture features to allow users to preserve and revise their TCFA artworks in 3D format. When users click the save button, the prototype generates prefabs for the selected vessel and the plant models that are in contact with the vessel's collider. This ensures that the relevant model data is saved and can be reloaded later. FloraJing also allows users to \rv{view their TCFA works} in reality \rv{(\textbf{DC5})} using the Quest 3's passthrough mode, as shown in Fig. \ref{fig:scene-panel-pass}(c).

\textbf{Considerations and Trade-offs of the Gravity System.}
In the design of the FloraJing, we chose not to implement a gravity system and instead opted for kinematic interactions with TCFA virtual materials. This decision was primarily driven by the feedback from users regarding the challenges they face in maintaining the stability of flower arrangements in real life (\textbf{DC6}). A gravity-based system could introduce additional complexities, such as flowers toppling over or requiring precise balancing, which might hinder creative expression and increase frustration. By removing this constraint, we aimed to provide a more fluid and intuitive experience, allowing users to focus on composition and aesthetics without being burdened by real-world structural concerns.

% \subsection{Pilot Testing and System Tuning}

% \sout{After developing the initial FloraJing prototype, we conducted a pilot testing phase to refine its features and optimize functionality for TCFA users. This phase involved a small group of participants (N=4), including two co-authors and two TCFA learners. Based on their feedback, we modified the menu picture proportions to create a more intuitive and visually balanced interface, and adjusted the lighting conditions to achieve a more comfortable environment.}

\section{USER STUDY}

\rv{The user study consisted of two parts to answer \textbf{RQ2}. First, we recruited 19 TCFA learners with diverse expertise to participate in a one-session TCFA practice, assessing whether the system could support TCFA practice. Then, we conducted a 7-day pilot study to evaluate whether and how FloraJing could sustain daily practice over time. All user studies were conducted in the same studio environment, with a floor space measuring 2.9$m$ $\times$ 3.9$m$. The space is clearly separated from other areas to avoid disturbances, ensuring that user experiments are conducted in a relatively quiet and safe environment. A set of Meta Quest 3 128GB was used to provide VR experience.}

\subsection{User Study 1: One TCFA Session Practice}
% \subsection{Participants}
\subsubsection{Participants}

Participants were recruited through social media, snowball sampling, and online communities. To ensure diversity in backgrounds and TCFA expertise, we collected information on their TCFA experience through a questionnaire, similar to the formative study, with an additional section on VR experience. Based on their responses, we recruited \rv{19} participants (\rv{10} male, \rv{9} female, average age = \rv{29.4}, SD = \rv{7.48}), as detailed in Table~\ref{table:Demographic background}. Participants received a \$15 compensation for their participation.

\subsubsection{Questionnaire}
The questionnaires includes three parts. They assess learners' TCFA \rv{practice outcomes}, \rv{system usability}, and \rv{relaxation state}. 

\rv{\textbf{TCFA Practice Questionnaire.}} To assess FloraJing's effectiveness in supporting TCFA practice, both in skill training and cultural knowledge acquisition, we selected \textit{Overall Practice Evaluation} along with two key TCFA skills: \textit{Composition} and \textit{Color Allocation} \cite{caiArtChineseFlower2017d, lianying_art_2020}, and two cultural knowledge aspects: \textit{Philosophical Aesthetics} and \textit{Cultural Context} \cite{xu_inheritance_2023, lianying_art_2020}. These formed the \textit{TCFA Practice Scale} (5-point Likert scale). For the complete list of questions, see Appendix Table \ref{app:one-session-practice}.

\rv{\textbf{Relaxation State Questionnaire.}} \label{Relaxation State} Relaxation state refers to a psychophysiological condition of low physical and psychological tension \cite{benson1974relaxation, 1999stress}, characterized by a decreased heart rate, lower metabolism, and slower breathing rate \cite{biofeedback}, accompanied by subjective feelings of calmness, refreshment, and a sense of timelessness \cite{ABCsmith2000abc}. The relaxation state was included in the user experience evaluation based on feedback from workshop participants who noted that practicing TCFA in real life provides a relaxing effect. Historically, TCFA was practiced by ancient Chinese as a means of relaxation and restoration \cite{li2002chinese}. Therefore, assessing participants' relaxation state before and after using FloraJing helps evaluate the system's ability to support TCFA practice.

\rv{To measure the user's relaxation state,} we referenced the State Trait Anxiety Inventory (STAI) \cite{spielberger1971state}, the Discrete Emotions Questionnaire (DEQ) \cite{discreteharmon2016} and the Relaxation State Questionnaire (RSQ) \cite{assessingsteghaus2022} and ultimately selected the following eight emotions to capture participants' emotional responses (Likert 7-point scale): \textit{Excited}, \textit{Tired}, \textit{Pleasant}, \textit{Nervous}, \textit{Relaxed}, \textit{Tense}, \textit{Calm} and \textit{Bored}. 
 % Modern research on flowers and flower arrangement has demonstrated their ability to enhance relaxation and calmness \cite{yang2022effects, liu2003physiological}, reduce fatigue M\cite{du_effect_2019}, and alleviate emotional stress \cite{lavin2021determining, morita2018increased}.

\rv{\textbf{System Usability \rv{Questionnaires}.}} \rv{Based on previous research on evaluating VR system usability \cite{huang2022factors, dayarathna2021assessment, lovasz2021usability}}, we adopted System Usability Scale (SUS, Likert 5-point scale) \cite{bangor2009determining} and NASA Task Load Index (NASA-TLX, Likert 5-point scale) \cite{hart1988development} to jointly assess FloraJing's usability, as shown in Appendix Table \ref{app:one-session-SUS-NASA}. SUS was used to measure the system's overall ease of use and user satisfaction, while NASA-TLX was to evaluate the perceived workload and stress during interaction with the system.
\begin{figure*}[t]
    \centering
    % \includesvg[width=0.95\linewidth]{Figures/flow chart.svg}
    \includegraphics[width=\linewidth]{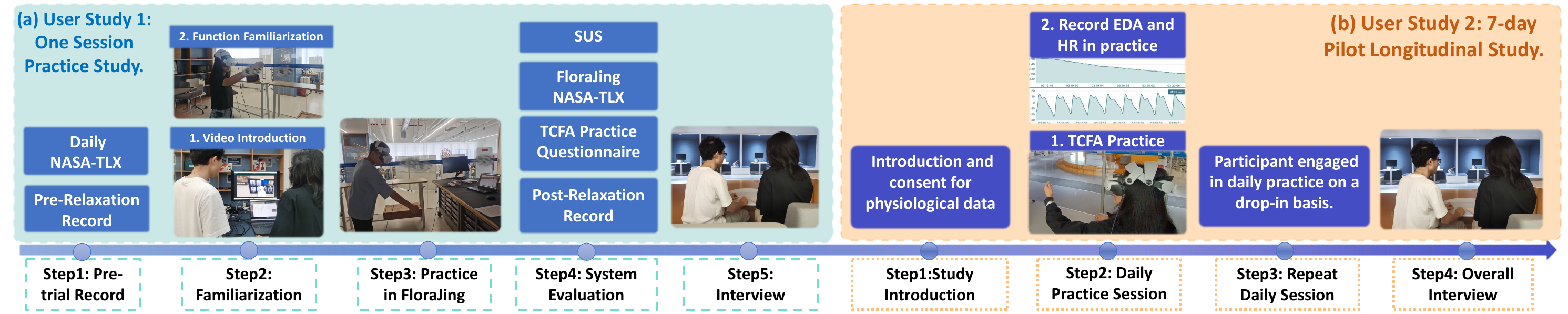}
    \caption{\textbf{\rv{The Study Procedure of One TCFA Session Practice and 7-day Pilot Longitude Study.}}}
    \label{fig: study procedure}  
    \Description{.}
\end{figure*}
\subsubsection{\rv{Procedure}}
The session comprised five parts, as shown in left part of Fig.~\ref{fig: study procedure}, and lasted approximately one hour.

1) \textbf{\textit{Introduction and Pre-\rv{Relaxation State} Record.}} Before starting, the participants were first briefed on the project and signed an informed consent form, which included permission for photos, screen, and audio recordings. Next, the participants filled out the NASA-TLX questionnaire (\textbf{Daily-NASA-TLX}) to assess their perceived workload during daily TCFA practice in reality. Following this, they completed a \textit{Relaxation State Questionnaire} \rv{(\textbf{Pre-Relax})}, as described in Section \ref{Relaxation State}, to gauge their \rv{relaxation state} before using the FloraJing.

2) \textbf{\textit{System Familiarization.}} Participants watched a two-minute video demonstrating FloraJing's basic operations, such as using pinch and poke gestures to interact with the menu, grabbing plants, and cutting materials with virtual scissors. They then wore the VR headset to explore the virtual environment until they self-reported having mastered the controls and interactions. Afterward, they removed the headset and took a three-minute break.

3) \textbf{\textit{FloraJing Experience.}} After the break, participants began their TCFA practice session in FloraJing, having ample time to create and save their arrangements until satisfied. The TCFA works created by the 19 participants are shown in Fig. \ref{fig:works}, with the pictures taken by them.

4) \textbf{\textit{Post-\rv{Relaxation State} Record and System Evaluation.}} After the experience session, participants were first asked to fill out a post-experiment \textit{Relaxation State Questionnaire} (\textbf{Post-Relax}). Following this, they completed the \textbf{SUS}, a NASA-TLX questionnaire (\textbf{VR-NASA-TLX}) to assess their perceived workload in the virtual environment, and a \rv{\textbf{TCFA Practice Questionnaire}} to evaluate the system's effectiveness in supporting their TCFA practice.

5) \textbf{\textit{Semi-Structured Interview.}} Finally, participants engaged in semi-structured interviews (about 30 minutes) to discuss their overall experience with FloraJing, the system's usability, the effectiveness of the skill practice features, and their suggestions or expectations for system improvement.

\subsection{\rv{User Study 2: A 7-day Pilot Longitudinal Study}}\label{Study2}

\subsubsection{\rv{Participants}}
\rv{To ensure continuity in the research and obtain in-depth feedback on TCFA practices, we recruited participants from User Study 1 to participate in User Study 2. Due to device limitations, Study 2 was conducted in the same lab space as User Study 1 on campus. To approximate a daily practice context, we recruited participants living on campus. Seven participants took part in User Study 2 over the 7-day period, as indicated by $*$ in Appendix Table \ref{table:Demographic background}.}

\subsubsection{\rv{Procedure}} The positive results regarding the subjective relaxation effects from participants in User Study 1 motivated us to incorporate biofeedback in User Study 2 to empirically asses the relaxation effects of FloraJing. We collected Electrodermal Activity (EDA) and Heart Rate (HR) which are well-established physiological measures used to assess emotional and stress responses in previous studies \cite{can2020relax,amore2019deep}. EDA reflects changes in skin conductance associated with sweat gland activity, which is often influenced by emotional arousal and relaxation states. HR, on the other hand, is indicative of autonomic nervous system activity, with lower rates typically associated with relaxation and higher rates linked to stress \cite{arabian2023analysis,caruelle2019use}. \rv{The procedure of the User Study 2 is shown in the right part of the Fig.\ref{fig: study procedure}.

\textbf{\textit{1) Introduction.}} Participants received an introduction and signed an informed consent form for data collection, including physiological data, videos, and photos. To align with the general daily practice of TCFA, we informed participants that the experimental space would be available with researcher assistance from 6:00 AM to 11:00 PM daily, and operated on a drop-in basis, allowing participants to join freely at their convenience.

\textbf{\textit{2) Daily Practice Session.}} Participants wore the Empatica E4 wristband on their non-dominant hand to record EDA and HR throughout each session\footnote{Empatica E4 wristband: www.empatica.com/en-int/research/e4/}. Once data stabilized, the TCFA practice began, with event tags marking the start and end. Each session, lasting 30 to 50 minutes, was recorded via video and screen capture. Over the 7-day period, participants completed 2 to 4 sessions. Specifically, three participants (P6, P11, P12) did two sessions, three (P1, P10, P16) did three, and one (P15) did four, totaling 19 practice sessions.

\textbf{\textit{3) Overall Semi-structure Interview.}}
At the end of User Study 2, participants took part in a semi-structured interview. Observations from User Study 1 showed that participants engaged in reflective thinking about their TCFA works, gained a better understanding of cultural meanings through practice, and expressed a desire to continue daily TCFA practice. This aligns with Kolb's experiential learning model, which highlights learning through a cycle of Concrete Experience, Reflective Observation, Abstract Conceptualization, and Active Experimentation \cite{kolb2014experiential}. Based on this, our interview questions focused on four areas: 1) Whether and how FloraJing helped improve their skills; 2) Whether and how participants engaged in reflection during VR practice; 3) How VR practice deepened their understanding of TCFA culture; 4) Whether they practiced TCFA in real life after VR sessions, and how VR practice influenced their real-life practice. \rv{For the complete list of questions, please refer to Appendix Table \ref{app:7-day-interview}.}
}

% daily practice session
% 随时来，drop-in
% screen recording
% overall interview

%These quantitative findings were then interpreted in the context of the qualitative insights from the interviews.
% \clearpage

%bio data, screen recoring, interview

\section{FINDINGS}
\begin{figure}[htbp]
  \newpage
  \centering
  \includegraphics[width=0.5\linewidth]{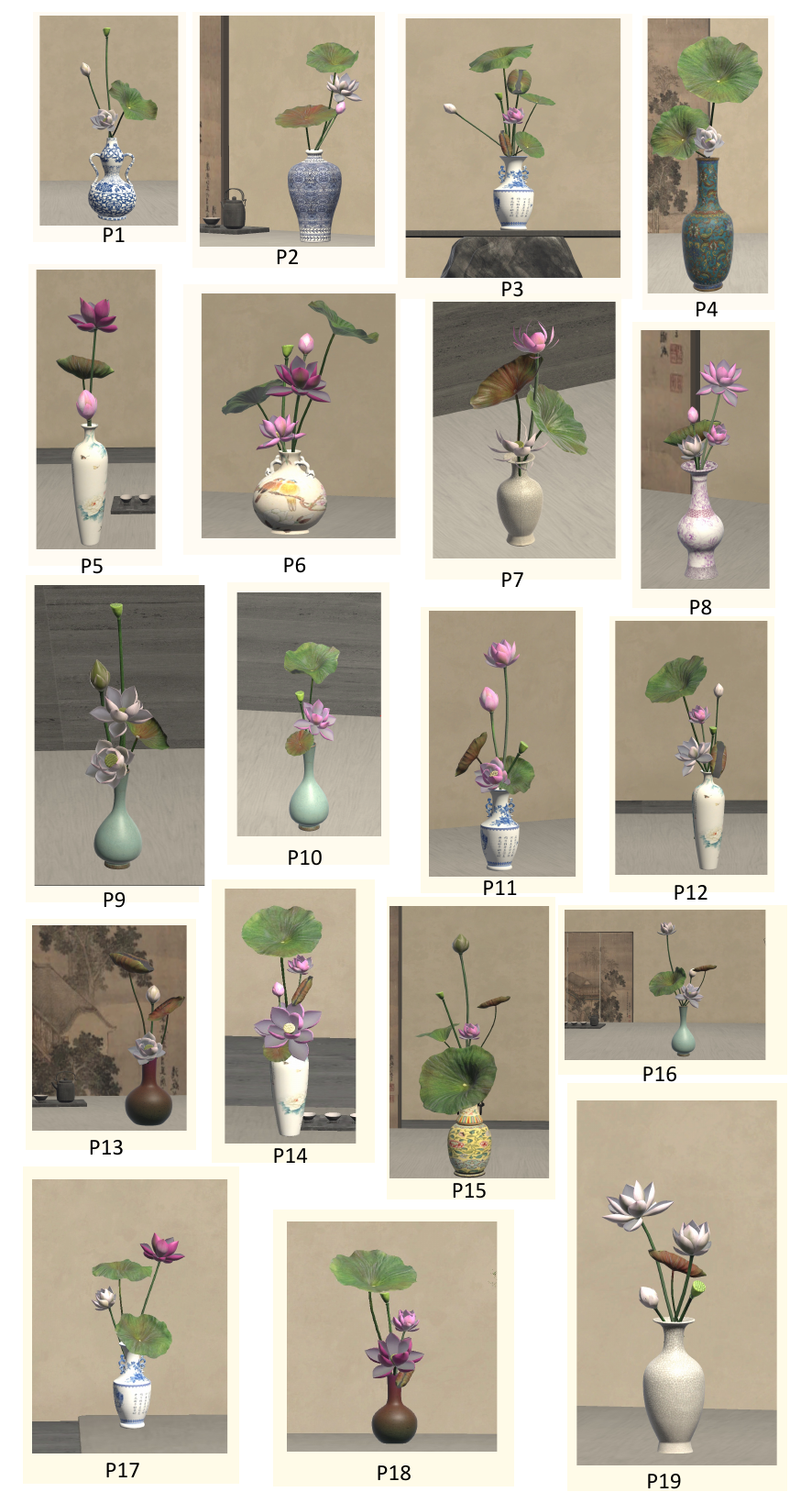}
  \caption{The TCFA works created and shot by the \rv{19} participants in User Study 1.}
  \label{fig:works}
  \Description{This figure shows the 19 traditional Chinese flower arrangement works created by the participants in our user study. Each work was captured in FLoraJing VR by the participant, containing virtual tea room environment and a virtual flower arrangement.}
\end{figure}
% \clearpage

\begin{figure*}[t]
  \centering
  \includegraphics[width=\linewidth]{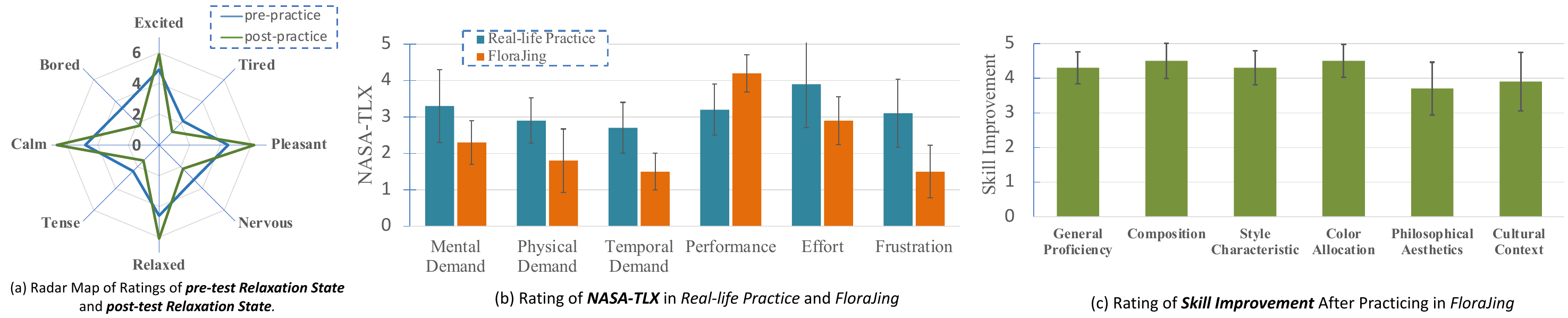}
  \caption{
  Results of the questionnaires in user study. \textbf{(a)} Radar Map of Ratings of \textit{pre-practice Relaxation State} and \textit{post-practice Relaxation State.} 
  \textbf{(b)} Bar Chart of Scores of NASA-TLX in \textit{Real-life Practice} and \textit{FloraJing}. 
  \textbf{(c)} Bar Chart of Scores of \textit{Skill Improvement} After Practicing in FloraJing.}
  \label{fig:emo-nasa-skill}
  \Description{
  Figure-a shows Radar Map of Ratings of pre-test Emotional State and post-test Emotional State.
  Figure-b shows the Bar Chart of Scores of NASA-TLX in Real-life Practice and FloraJing.
  Figure-c shows the Bar Chart of Scores of Skill Improvement After Practicing in FloraJing. 
  }
\end{figure*}

\subsection{Overall Experience.} 
Overall, all participants indicated that the experience of practicing TCFA in FloraJing was novel, interesting, immersive, convenient and \rv{low-cost}. Every participant reported experiencing the same flow state \cite{flow} in VR as they did during real-life TCFA practice. They felt that their skills and \rv{cultural understanding} of TCFA were enhanced through the practice in FloraJing, and they expressed a willingness to use it for daily TCFA practice in their lives in the future.

\subsection{FloraJing's Usability}
The participants' feedback on the prototype's usability was gathered from the data collected in User Study 1, including the \textit{SUS}, \textit{Daily-NASA-TLX}, and \textit{VR-NASA-TLX} questionnaires, as well as the screen recording time spent by users on system familiarization and practice complexation.

\textbf{Reduction in Time and Physical Load.} Compared to the time participants reported spending on daily TCFA practice \rv{(Daily: $M = 82.5 min$, $Max = 210 min$, $Min = 42 min$)}, all participants expressed that FloraJing \rv{(VR: $M = 15.6 min$, $Max = 35 min$, $Min = 9 min$)} significantly saved time by eliminating the need to purchase flowers, stabilize arrangements, and clean up the environment. We also compared the \textit{Daily-NASA-TLX} and \textit{VR-NASA-TLX} scores (as shown in Fig.~\ref{fig:emo-nasa-skill}(b)). A Shapiro-Wilk test on these two NASA-TLX scores indicated that the data is not normally distributed ($p < 0.05$). Therefore, Mann-Whitney U tests were performed. The results show that, compared to real-life practice, practicing TCFA in FloraJing significantly reduced physical, mental, and temporal demands \rv{(each $p < 0.001$)}.

\textbf{Ease of Use.} The \textit{SUS} score analysis \cite{bangor2009determining} indicates that FloraJing's various functions are well-integrated, easy to learn (\rv{$M = 4.7, SD = 0.71$}), easy to operate (\rv{$M = 4.2, SD = 0.76$}), and smooth in use (\rv{$M = 3.8, SD = 0.82$}). The short time participants spent during the system familiarization phase ($M = 6 min, Max = 9 min, Min = 4 min$) also suggests that the system is intuitive and easy to learn.

\subsection{Effectiveness of FloraJing supporting TCFA Daily practice}

Participants from \textit{User Study 1} recognized FloraJing's contribution to enhancing their overall TCFA skills \rv{($M = 4.2, SD = 0.52$)}. Beginners reported the greatest learning benefits, though no significant difference was observed across skill levels (p > 0.05). \rv{Most participants (N=12) appreciated the diverse selection of virtual materials, which provided greater creative flexibility than in reality.}

\begin{figure*}[htb]
    \centering
    \includegraphics[width=0.9\linewidth]{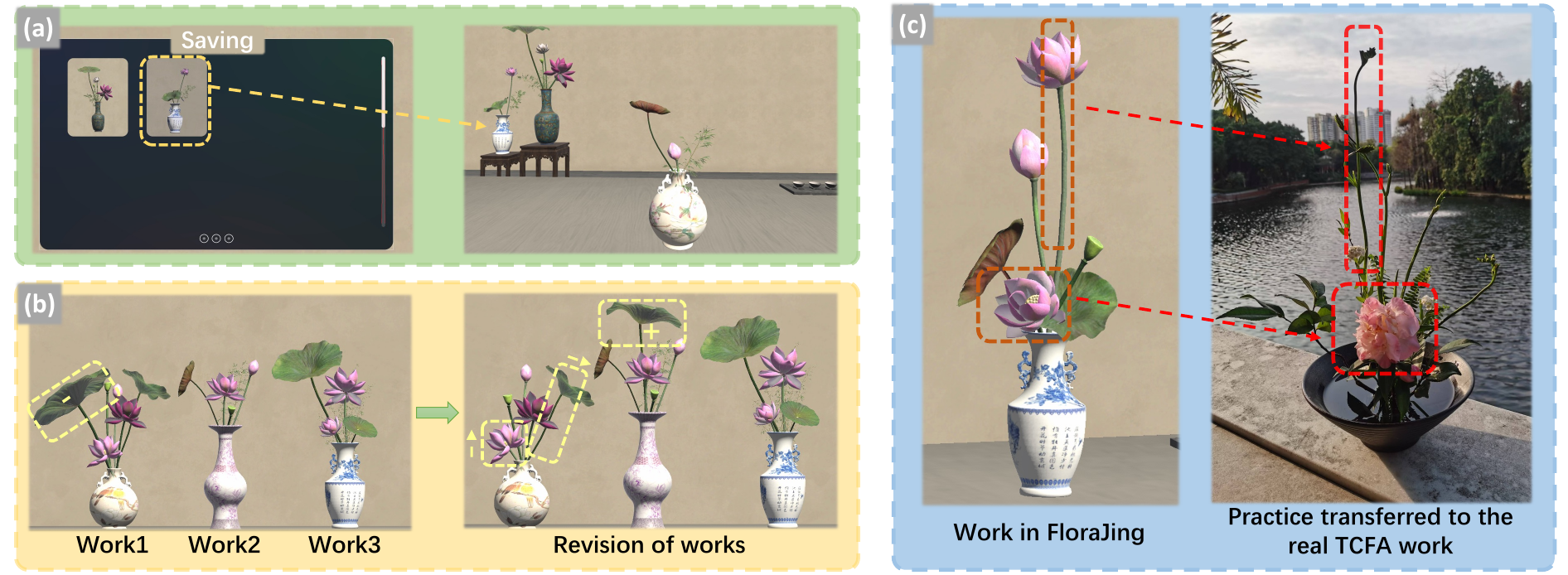}
    \caption{\rv{\textbf{(a)} In User Study 2, participants could review their previous TCFA works while creating new ones.
    \textbf{(b)} The series of works by P6 during the 7-day practice, with her reflections and adjustments to previous creations.
    \textbf{(c)} P11 transferred her virtual TCFA work in VR into a physical arrangement.}}
    \label{fig: reflection}  
    \Description{.}
\end{figure*}

\rv{\subsubsection{Facilitate Progressive Reflection and Improvement Based on Previous TCFA Work Records.}
All participants from the 7-day pilot study indicated that the ability to view and modify previous TCFA works in FloraJing significantly facilitated their continued reflection and improvement, which was evident in their exploration of different styles, refinement of past works, and increased motivation to learn more about TCFA knowledge and skills.We observed that all participants began their practice in FloraJing by reviewing their previous works, as shown in Fig. ~\ref{fig: reflection}(a). These works served as references, inspiring them to experiment with different styles, such as alternative color allocation and composition, to challenge themselves and create new, distinct pieces. For example, P15 frequently revisited her earlier creations while selecting vases and flowers to differentiate her new work from the previous ones.

After completing one session, most participants (N=5) laid out all their works for extended observation, and comparison, which further led to reflection and modification. For instance, after completing her third arrangement, P6 believed that it had the most harmonious proportions. In contrast, she identified issues in her earlier works, \textit{``The first was top-heavy with a crowded base, lacking the beauty of blank space in TCFA, and the second suffered from disharmony between the flowers and the vase proportions.”} Drawing from insights gained through comparative observation, she revisited and refined her earlier works, as shown in Fig. ~\ref{fig: reflection}(b). P12, on the other hand, was surprised by how her aesthetic preferences constrained her creativity. After her fourth practice, she realized that despite attempts to create something completely different, the resulting structures across her works were similar, with only the colors changing. Reflecting on this, P12 remarked, \textit{``I might like this style, but it also shows my lack of learning. I need to explore more works and different expressions to enrich myself.”}}

\subsubsection{Transfer Virtual Creative Experience to Real-Life TCFA Practice.}
Four participants from Study 2 practiced TCFA in real life during the experiment. They indicated that their practice in FloraJing motivated them to engage in TCFA practice in real life, acknowledging that VR practice helped them better conceptualize their arrangements, while also boosting their confidence and boldness to experiment with a broader range of materials and attempt more complex compositions and color combinations.

Through the passthrough mode, they could test the alignment of their virtual works with reality, inspiring them to recreate similar arrangements. And when encountering similar flowers, vases, or styles in real life, they would recall their VR creations as valuable references to aid in pre-arrangement conceptualization. For instance, P11 used her VR TCFA work as inspiration, adapting it with different materials to replicate similar structures and colors in a real-life arrangement, as shown in Fig. ~\ref{fig: reflection}(c). Additionally, participants mentioned that shorter VR practice sessions allowed them to create a wider variety of samples for future reference. 

\subsubsection{Increase understanding of TCFA within immersive cultural-enriched Environment}
Participants from both user studies indicated that the combination of traditional Chinese-style sound and visuals in FloraJing created a culturally immersive atmosphere, helping them quickly transition from their everyday environment to engage in TCFA as a cultural activity. This cultural-enriched virtual environment deepened their understanding of the \textit{Philosophical Aesthetics} ($M = \rv{3.9}, SD = \rv{0.73}$) and \textit{Cultural Context} ($M = \rv{3.9}, SD = \rv{0.84}$) of TCFA.  For example, P13 mentioned being inspired by the virtual Zen-style tea room, which led her to leave more blank spaces in her arrangement to evoke a sense of breathability and potential. P11 recognized the background music, a traditional Chinese piece named High Mountains Flowing Water, and the Chinese landscape painting screen, which inspired him to showcase the natural growth and blossoming of the lotus flowers in the wild. 

Additionally, the TCFA materials prompted participants to associate their creations with related literary works (N=4). For instance, P1 recalled the classical Chinese poetry \textit{Ode to the Lotus}, which inspired him to emphasize the noble qualities of the lotus in his arrangement. The TCFA works created by participants from User Study 1 are shown in Fig. ~\ref{fig:works}.

\subsection{Relaxation State Changes} 

\subsubsection{Subjective Relaxation Effect.}
The radar chart Fig. \ref{fig:emo-nasa-skill}(a) illustrates participants' \rv{subjective
relaxation state} before and after using the system. A Shapiro-Wilk test indicated that the data of the \textit{Relaxation State Questionnaire} is normally distributed (p > 0.05). Given the pre- and post-test design within subjects, we performed paired t-tests to assess significant differences in emotional ratings. The analysis revealed significant effects on all dimensions. All participants expressed that while arranging flowers in FloraJing VR, the natural, realistic, and coherent interactions allowed them to enter a flow state similar to that of real-life flower arrangement. \rv{Due to the immersive nature of VR, which isolates external distractions and creates a cultural atmosphere through audiovisual elements, some participants (N=7) indicated that practicing flower arrangement in VR provided a greater relaxation effect than real-life practice.} P13 shared, \textit{``Although I had sweat on my head after taking off the HMD, I didn't feel any fatigue. I could clearly feel that my mind and body were in a joyful state, and this feeling was even stronger than when arranging flowers in real life. I actually worked a night shift yesterday and then an early shift today, so I was a bit tired. But after doing this today, I felt relaxed and awake. It's quite amazing.''}

\subsubsection{\rv{EDA Data Indicates Relaxation Effect}}
% overall Trend
We utilized relative change to assess both EDA and HR during the TCFA practice \cite{li2022electrodermal, gaertner2023relaxing}. The difference between the average EDA and HR values during the activity and their respective baseline values was calculated. A relaxation state for EDA was defined as a relative decrease exceeding 10\%, and for HR, a decrease exceeding 5\%. In terms of results, based on EDA data, seven users entered a relaxation state after the daily practice, with 15 sessions (78.9\%) out of 19 showing this effect. For HR data, five users reached a relaxation state in 8 sessions (42.1\%).

\subsection{Expectations for Improvement}

\subsubsection{Enhancing Cultural Immersion through Olfactory Simulation of Traditional Chinese Incense}
While participants appreciated the visual and auditory simulations in FloraJing, they pointed out the absence of olfactory enhancement. Interestingly, even though floral scents are integral to the flower arrangement experience in reality, most participants (5 out of 7 who mentioned the lack of olfactory input) expressed a preference for the simulation of traditional Chinese incense rather than floral scents. When further asked why they preferred Chinese incense over floral scents, participants explained that rather than pursuing sensory realism, they more valued a multisensory experience that enhances the cultural atmosphere and immersion. As P6 said, \textit{``No, I didn't think of flower scents. Because it's Chinese flower arrangement, I'd prefer a deeper immersion in Chinese aesthetics.''}

\subsubsection{Personalized Feedback on TCFA Artwork}
Participants had personalized opinions on receiving feedback after practicing in FloraJing. Most (N=10) wanted feedback, mainly from instructors, while others saw practice as a form of self-relaxation, fearing feedback might add stress. To ease this, P4 and P3 suggested an AI instructor. For those who wanted feedback, their preferences varied. Some (N=4) wished to see the entire process from a first-person perspective of how the teacher freely modified their arrangement, along with the final result. Some (N=2) preferred to see only the final result, choosing to reflect on and appreciate the changes.

\subsubsection{Different Requirements for System Realism}
Participants expressed varying requirements for system realism. Regarding the replacement of gravity with kinematic interaction, participants at different learning stages had distinct experiences and needs. All beginners and most competent participants (N=9) found the absence of gravity beneficial, making the process more enjoyable and allowing them to focus on the arrangement without real-world constraints. For example, this study involved vase arrangements, which are more challenging and typically introduced in later TCFA learning stages. Unlike arrangements in open floral vessels, vase arrangements require stabilizing stems within a nearly invisible interior space, making balance difficult. In FloraJing, many beginners had their first opportunity to explore and appreciate vase arrangements. However, participants with higher realism expectations (N=4), mainly experts, felt that the lack of gravity simulation fell short of their standards. They acknowledged the system's value for extensive daily practice and skill training but expected more realistic physics simulation to support high-level artistic creation.

As for feedback on real-world tactile sensations, preferences did not differ by TCFA skill level but were driven by different needs. Some participants (N=6) wanted a stick-like tool to support their fingers and reduce effort, while others (N=4) preferred dynamic feedback that matched the virtual model for a more realistic flower-arranging experience.

\section{DISCUSSION AND FUTURE WORK}
% 2.Some design decisions are questionable, such as replacing gravity with kinematic interaction. As a developer, I understand this choice, but it has significant implications for learning the skill, which is one of the purposes of daily practice, as I interpret it. This raises questions about whether the study's understand the purpose of daily practice in the art, or learning skills, or if it is merely an exploration of user requirements without a solid understanding of the purpose of practice. ---multi-model/different learning level

% Kolb's experiential learning model, which emphasizes learning through a cycle of Concrete Experience, Reflective Observation, Abstract Conceptualization, and Active Experimentation \cite{kolb2014experiential}. 

% This expected pathway aligns with Kolb's experiential learning model, which emphasizes learning through a cycle of Concrete Experience, Reflective Observation, Abstract Conceptualization, and Active Experimentation \cite{kolb2014experiential}.

% However, compared to technical skill training, ICH practice
% places greater emphasis on fostering interest, understanding, and creativity within the concerned communities [6].
% Achieving this goal requires reflecting on concrete ICH practices, using methods such as active imagination to abstract
% cultural meanings from specific technical rules [2], and continuing to actively engage in ICH practices.

Informed by a formative study with TCFA learners of varied expertise, we identified six DCs and developed FloraJing, a VR application that provides rich cultural environments and accessible materials and tools. \rv{We then conducted a one-session TCFA practice with FloraJing to assess the system's support for TCFA practice, followed by a 7-day pilot longitudinal study to evaluate its effectiveness in sustaining TCFA daily practice. The results demonstrated that FloraJing facilitates TCFA learners' progressive reflection, skill improvement and enhances cultural understanding of TCFA. Participants also expressed diverse needs regarding the system's realism and the types of feedback they received on their creations. In this section, we discuss how our findings inform the design of systems that support daily practice for ICH with VR.}

\subsection{Supporting ICH Daily Skill Practice through VR} 
\rv{In this section, we refer to Kolb's experiential learning model \cite{kolb2014experiential}, which aligns with the learning patterns of participant' practices, to propose design implications for future research.}

\subsubsection{Enhancing \rv{Concrete} Creative Experience by Reducing Peripheral Tasks in ICH Practice}
\rv{In Kolb's experiential learning model, the \textit{Concrete Experience} phase refers to learners engaging directly with a related hands-on experience} \cite{kolb2014experiential}. Our findings indicate that activities such as buying flowers and organizing the workspace, which are external to the creative process, significantly hinder TCFA learners from maintaining regular daily practice and greatly impact their creative experience.

For many ICH, lengthy material preparation is required. For example, traditional Chinese lacquerware involves manually extracting raw lacquer sap from trees, followed by extensive processing.Materials once commonly available for daily practice have become rare, making preparation time-consuming and challenging. This forces practitioners to spend disproportionate time sourcing materials rather than focusing on artistic mastery. Additionally, the fast pace of modern life makes it difficult to sustain ICH practices that require prolonged daily training. VR can help integrate ICH practice into contemporary daily life by reducing peripheral tasks that may require scarce physical resources \cite{ArtPractice, som2023virtual} while preserving cultural and artistic experiences. In this study, participants noted that the high-fidelity TCFA virtual materials and convenient cleaning features in FloraJing allowed them to focus more deeply on the creative process, enhancing immersion and enabling full attention to artistic expression. This also shortened practice time, making daily engagement more flexible and convenient.
To maximize the benefits of VR in making daily practice more flexible for ICH learners, we propose \textbf{Design Implication 1: Enabling ICH learners to focus more on their artistic expression by reducing peripheral activities and disruptions in their daily practice.}

% Studies have demonstrated that having a well-organized and prepared environment reduces cognitive load for creators, while disruptions in these stages can slow down the creative process and lead to frustration or reduced creative satisfaction \cite{Creativeflow, Creativeprocesses}. 
\subsubsection{Using Multimodal Co-creation of ICH Cultural Atmosphere to \rv{Enhance Understanding of Abstract Aesthetic Concepts}}The \textit{Abstract Conceptualization} phase in Kolb's experiential learning model involves learners forming deeper insights and frameworks by interpreting concrete experiences. However, unlike technical skill training, ICH practice focuses more on fostering cultural and aesthetic understanding within communities \cite{bakar2011intangible}, such as the philosophical wisdom developed over generations. These intangible abstract concepts stem from the cultural contexts that shape their meaning \cite{brown2005heritage}. However, in modern industrial and information societies, such contexts are no longer part of daily life. VR can support this process by reconstructing or even enhancing cultural contexts, activating learners' imagination, and further inspiring their own interpretations during practice. Participants' feedback in our study indicated that the visual elements (e.g., the Chinese tea room and TCFA flowers) and auditory simulations (e.g., traditional Chinese music) together shaped a cultural atmosphere that helped them conceptualize TCFA's aesthetic principles, such as \textit{intended blank} and the \textit{harmony between humans and nature}, and facilitated associations with related literary knowledge.

When participants expressed a desire for enhanced sensory experiences, they preferred the simulation of traditional Chinese incense over floral fragrances, even though floral scents are naturally part of real-world flower arranging. While sensory realism is often emphasized in many VR skill practice systems in pursuit of better skill training outcomes \cite{motorskill1, knottyingskills}, our results show that, instead of pursuing sensory fidelity, ICH learners favor a richer cultural atmosphere and immersion. This approach helps them better understand and engage in the ICH aesthetic experience. To create a more culturally immersive experience in VR systems for ICH learners' daily practice, we propose \textbf{Design Implication 2: Enable ICH learners to better understand and immerse themselves in the cultural atmosphere of ICH through multimodal co-creation aligned with its cultural essence.}

\subsubsection{Enhancing Reflection with Progressive ICH Work Records}
In Kolb's experiential learning model, the \textit{Reflective Observation} phase involves learners contemplating their experiences and drawing insights to guide future actions \cite{kolb2014experiential}. Research on creative skill improvement, particularly in art and design education, indicates that keeping records of past works can substantially impact skill development in creative practices \cite{samaniego2024creative, papachristopoulos2023positive}. However, some traditional ICH practices like TCFA face intrinsic limitations: physical artifacts (e.g., flowers) are irreversible and difficult to preserve once completed. This hinders iterative refinement, which is a critical component of the learning process.

VR circumvents these constraints by digitizing the entire creative lifecycle in 3D format. Practicing in VR enables all ICH works to be saved, fostering reflection by allowing learners to observe and modify past creations without limitations. These previous works serve as valuable input for new ones, while accessible display and selection mechanisms further stimulate synergistic effects in progressive ICH practice. Our findings suggest that the ability to review and modify previous TCFA works in FloraJing significantly supported learners' reflection and improvement. Participants in the 7-day pilot study made cross-temporal comparisons by visually overlaying past and current works to track self evolution. They emphasized that revisiting and adjusting past works was crucial for personal growth, helping them recognize strengths and weaknesses in their arrangements. This process fostered a deeper understanding of their ability and encouraged continuous refinement. Based on these insights, we propose \textbf{Design Implication 3: Facilitating progressive reflection and skill improvement by enabling learners to revisit and modify previous ICH works in virtual systems, enhancing their learning through continuous observation and refinement.}

\subsection{Enhancing Emotional Well-being in ICH Pratices Through VR}
Previous studies indicate that engaging with ICH activities \rv{in VR can help ICH learners relax}. The immersive experience of ICH activities (e.g., traditional crafts) fosters a sense of connection with local culture and contributes to well-being, stress relief, and personal happiness \cite{sayer2015can, ning2023analysis, yang2023impact}. However, in real-life scenarios, environmental distractions and peripheral tasks often interrupt the immersive experience of ICH \cite{Creativeflow, Creativeprocesses}, limiting the emotional benefits and flow states that arise from sustained attention and mindfulness.

Our findings suggest that VR can help users overcome these challenges by providing a culturally rich virtual environment. In VR-based ICH practices, users can quickly immerse themselves, achieving or even surpassing the flow experiences found in real-life practice. They report feeling both relaxed and energized, deeply focused yet not fatigued, and even experiencing a sense of self-restoration \cite{kohut2009restoration}. VR enhances focus by isolating users from external distractions, inducing a flow state, and improving concentration \cite{Transport}. This benefit is particularly valuable for activities requiring sustained attention or mindfulness \cite{zenvr_2022, mindfulnesspractice}, such as ICH practices. Given VR's potential to create immersive, culturally enriched environments that foster deeper emotional engagement, we advocate for future research to further explore its role in enhancing emotional well-being in ICH practices.

\subsection{Limitation and Future Work}

\subsubsection{Multimodal Simulation for Enhanced Realism in VR ICH Practice}
While FloraJing serves as a complementary practice tool with visual and interactive fidelity, it currently lacks tactile feedback (e.g., stem textures) and olfactory cues (e.g., floral scents), which are critical to TCFA's connection with nature. Authentic sensory experiences may help establish bonds between users and ICH practices, and meet the artistic demands of advanced practitioners. Future work could explore integrating multimodal simulations while maintaining convenience, such as hybrid AR-VR systems that blend virtual and physical materials, to create a more immersive and realistic practice experience.

\subsubsection{\rv{Lab-based Longitudinal Study}}
\rv{Influenced by the constraints of available equipment, our 7-day Pilot Longitudinal Study is the setting in a campus lab environment. Although we selected participants who lived on campus and extended the available time as much as possible,  The controlled lab environment might not fully capture the diversity of real-life settings. Future work could deploy the study in participants' homes for a longer period.}

\subsubsection{\rv{Small Sample Size in TCFA Workshop}}
\rv{One limitation of this study is the small sample size in the initial workshop, which might affect the generalizability of the findings regarding the challenges faced by TCFA learners. While the small sample allowed us to gain valuable insights into the specific difficulties encountered by a diverse group of TCFA learners, we recognize that a larger sample size could provide more comprehensive data and a better understanding of these challenges across various skill levels.}

\subsubsection{\rv{Control Group for Comparison}} \rv{This study does not include a control group, as the primary goal of this exploratory research was to investigate how VR features could facilitate ICH daily practice, focusing on the design process and rationale. However, future work could compare the effectiveness of VR with other technologies (e.g., AR) in supporting ICH daily practice to gain different insights into the creative practice of ICH in a contemporary technological era.}

\subsubsection{Possibility of Expert Involvement}
A limitation of this study is that two of the researchers were still in the learning phase of TCFA. While expert participants provided valuable support through their professional knowledge and guidance on venue setup, their level of involvement was not fully integrated into the research process, such as informing the data interpretation. In future work, we suggest involving ICH experts more deeply, potentially as co-authors or conducting expert evaluations for more critical insights.

% \section{LIMITATION AND }
% \input{Chapters/08Limitation} 

\section{CONCLUSION}
\rv{This study highlights the potential of VR as an effective tool for facilitating ICH daily practice through a case study of TCFA.} We conducted three workshops to identify specific challenges TCFA learners encounter and derived six DCs. Based on these considerations, we proposed FloraJing, a VR application that provides rich cultural environments and accessible materials and tools for the daily practice of TCFA. We conducted a one-session TCFA practice with FloraJing to assess the system's support for TCFA practice, followed by a 7-day pilot longitudinal study to evaluate its effectiveness in sustaining TCFA daily practice. The results demonstrated that FloraJing facilitates TCFA learners' progressive reflection, skill improvement, and enhances cultural understanding of TCFA. Our work provides a foundation for future research in this field, highlighting the potential of VR applications designed for ICH daily practice, \rv{both in knowledge and skills.}

\begin{acks}
\begin{sloppypar}
This work is partially supported by the Guangzhou-HKUST(GZ) Joint Funding Project (No. 2024A03J0617),  Education Bureau of Guangzhou Municipality Funding Project (No. 2024312152), Guangzhou Higher Education Teaching Quality and Teaching Reform Project (No. 2024YBJG070), Guangdong Provincial Key Lab of Integrated Communication, Sensing and Computation for Ubiquitous Internet of Things (No. 2023B1212010007), the Project of DEGP (No.2023KCXTD042), and the Guangzhou Science and Technology Program City-University Joint Funding Project (No. 2023A03J0001).
\end{sloppypar}
\end{acks}

\bibliographystyle{ACM-Reference-Format}
\bibliography{reference}


\begin{thebibliography}{89}

%%% the Web address.

\ifx \showCODEN    \undefined \def \showCODEN     #1{\unskip}     \fi
\ifx \showDOI      \undefined \def \showDOI       #1{#1}\fi
\ifx \showISBNx    \undefined \def \showISBNx     #1{\unskip}     \fi
\ifx \showISBNxiii \undefined \def \showISBNxiii  #1{\unskip}     \fi
\ifx \showISSN     \undefined \def \showISSN      #1{\unskip}     \fi
\ifx \showLCCN     \undefined \def \showLCCN      #1{\unskip}     \fi
\ifx \shownote     \undefined \def \shownote      #1{#1}          \fi
\ifx \showarticletitle \undefined \def \showarticletitle #1{#1}   \fi
\ifx \showURL      \undefined \def \showURL       {\relax}        \fi
% invisible to TeX
\providecommand\bibfield[2]{#2}
\providecommand\bibinfo[2]{#2}
\providecommand\natexlab[1]{#1}
\providecommand\showeprint[2][]{arXiv:#2}

\bibitem[Alves(2018)]%
        {activeimaginationunderstanding}
\bibfield{author}{\bibinfo{person}{Susana Alves}.} \bibinfo{year}{2018}\natexlab{}.
\newblock \showarticletitle{Understanding intangible aspects of cultural heritage: The role of active imagination}.
\newblock \bibinfo{journal}{\emph{The Historic Environment: Policy \& Practice}} \bibinfo{volume}{9}, \bibinfo{number}{3-4} (\bibinfo{year}{2018}), \bibinfo{pages}{207--228}.
\newblock


\bibitem[Anastasovitis and Roumeliotis(2020)]%
        {2020creative}
\bibfield{author}{\bibinfo{person}{Eleftherios Anastasovitis} {and} \bibinfo{person}{Manos Roumeliotis}.} \bibinfo{year}{2020}\natexlab{}.
\newblock \showarticletitle{Creative Industries and Immersive Technologies for Training, Understanding and Communication in Cultural Heritage}. In \bibinfo{booktitle}{\emph{Euro-Mediterranean Conference}}. \bibinfo{publisher}{Springer}, \bibinfo{pages}{450--461}.
\newblock


\bibitem[Anders~Ericsson(2008)]%
        {deliberate}
\bibfield{author}{\bibinfo{person}{K Anders~Ericsson}.} \bibinfo{year}{2008}\natexlab{}.
\newblock \showarticletitle{Deliberate practice and acquisition of expert performance: a general overview}.
\newblock \bibinfo{journal}{\emph{Academic emergency medicine}} \bibinfo{volume}{15}, \bibinfo{number}{11} (\bibinfo{year}{2008}), \bibinfo{pages}{988--994}.
\newblock


\bibitem[Arabian et~al\mbox{.}(2023)]%
        {arabian2023analysis}
\bibfield{author}{\bibinfo{person}{Herag Arabian}, \bibinfo{person}{Ramona Schmid}, \bibinfo{person}{Verena Wagner-Hartl}, {and} \bibinfo{person}{Knut Moeller}.} \bibinfo{year}{2023}\natexlab{}.
\newblock \showarticletitle{Analysis of EDA and Heart Rate Signals for Emotional Stimuli Responses}. In \bibinfo{booktitle}{\emph{Current Directions in Biomedical Engineering}}, Vol.~\bibinfo{volume}{9}. De Gruyter, \bibinfo{pages}{150--153}.
\newblock


\bibitem[Bakar et~al\mbox{.}(2011)]%
        {bakar2011intangible}
\bibfield{author}{\bibinfo{person}{Aisyah~Abu Bakar}, \bibinfo{person}{Mariana~Mohamed Osman}, {and} \bibinfo{person}{Syariah Bachok}.} \bibinfo{year}{2011}\natexlab{}.
\newblock \showarticletitle{Intangible Cultural Heritage: Understanding and Manifestation}. In \bibinfo{booktitle}{\emph{International Conference on Universal Design in Built Environment}}, Vol.~\bibinfo{volume}{22}. \bibinfo{pages}{23}.
\newblock


\bibitem[Bangor et~al\mbox{.}(2009)]%
        {bangor2009determining}
\bibfield{author}{\bibinfo{person}{Aaron Bangor}, \bibinfo{person}{Philip Kortum}, {and} \bibinfo{person}{James Miller}.} \bibinfo{year}{2009}\natexlab{}.
\newblock \showarticletitle{Determining what individual SUS scores mean: Adding an adjective rating scale}.
\newblock \bibinfo{journal}{\emph{Journal of usability studies}} \bibinfo{volume}{4}, \bibinfo{number}{3} (\bibinfo{year}{2009}), \bibinfo{pages}{114--123}.
\newblock


\bibitem[Benson et~al\mbox{.}(1974)]%
        {benson1974relaxation}
\bibfield{author}{\bibinfo{person}{Herbert Benson}, \bibinfo{person}{John~F Beary}, {and} \bibinfo{person}{Mark~P Carol}.} \bibinfo{year}{1974}\natexlab{}.
\newblock \showarticletitle{The relaxation response}.
\newblock \bibinfo{journal}{\emph{Psychiatry}} \bibinfo{volume}{37}, \bibinfo{number}{1} (\bibinfo{year}{1974}), \bibinfo{pages}{37--46}.
\newblock


\bibitem[Botella and Lubart(2016)]%
        {Creativeprocesses}
\bibfield{author}{\bibinfo{person}{Marion Botella} {and} \bibinfo{person}{Todd Lubart}.} \bibinfo{year}{2016}\natexlab{}.
\newblock \showarticletitle{Creative processes: Art, design and science}.
\newblock \bibinfo{journal}{\emph{Multidisciplinary contributions to the science of creative thinking}} (\bibinfo{year}{2016}), \bibinfo{pages}{53--65}.
\newblock


\bibitem[Brown(2005)]%
        {brown2005heritage}
\bibfield{author}{\bibinfo{person}{Michael~F Brown}.} \bibinfo{year}{2005}\natexlab{}.
\newblock \showarticletitle{Heritage trouble: recent work on the protection of intangible cultural property}.
\newblock \bibinfo{journal}{\emph{International Journal of Cultural Property}} \bibinfo{volume}{12}, \bibinfo{number}{1} (\bibinfo{year}{2005}), \bibinfo{pages}{40--61}.
\newblock


\bibitem[Cai(2017)]%
        {caiArtChineseFlower2017d}
\bibfield{author}{\bibinfo{person}{Zhongjuan Cai}.} \bibinfo{year}{2017}\natexlab{}.
\newblock \bibinfo{booktitle}{\emph{Art of {{Chinese Flower Arrangement}}} (\bibinfo{edition}{1st} ed.)}.
\newblock \bibinfo{publisher}{Shanghai Press}.
\newblock
\showISBNx{978-1-60220-026-5}


\bibitem[{\c{C}}akiro{\u{g}}lu and G{\"o}ko{\u{g}}lu(2019)]%
        {ccakirouglu2019development}
\bibfield{author}{\bibinfo{person}{{\"U}nal {\c{C}}akiro{\u{g}}lu} {and} \bibinfo{person}{Seyfullah G{\"o}ko{\u{g}}lu}.} \bibinfo{year}{2019}\natexlab{}.
\newblock \showarticletitle{Development of fire safety behavioral skills via virtual reality}.
\newblock \bibinfo{journal}{\emph{Computers \& Education}}  \bibinfo{volume}{133} (\bibinfo{year}{2019}), \bibinfo{pages}{56--68}.
\newblock


\bibitem[Can et~al\mbox{.}(2020)]%
        {can2020relax}
\bibfield{author}{\bibinfo{person}{Yekta~Said Can}, \bibinfo{person}{Heather Iles-Smith}, \bibinfo{person}{Niaz Chalabianloo}, \bibinfo{person}{Deniz Ekiz}, \bibinfo{person}{Javier Fern{\'a}ndez-{\'A}lvarez}, \bibinfo{person}{Claudia Repetto}, \bibinfo{person}{Giuseppe Riva}, {and} \bibinfo{person}{Cem Ersoy}.} \bibinfo{year}{2020}\natexlab{}.
\newblock \showarticletitle{How to relax in stressful situations: a smart stress reduction system}. In \bibinfo{booktitle}{\emph{Healthcare}}. MDPI, \bibinfo{pages}{100}.
\newblock


\bibitem[Caruelle et~al\mbox{.}(2019)]%
        {caruelle2019use}
\bibfield{author}{\bibinfo{person}{Delphine Caruelle}, \bibinfo{person}{Anders Gustafsson}, \bibinfo{person}{Poja Shams}, {and} \bibinfo{person}{Line Lervik-Olsen}.} \bibinfo{year}{2019}\natexlab{}.
\newblock \showarticletitle{The use of electrodermal activity (EDA) measurement to understand consumer emotions--A literature review and a call for action}.
\newblock \bibinfo{journal}{\emph{Journal of Business Research}}  \bibinfo{volume}{104} (\bibinfo{year}{2019}), \bibinfo{pages}{146--160}.
\newblock


\bibitem[Chemi(2016)]%
        {Creativeflow}
\bibfield{author}{\bibinfo{person}{Tatiana Chemi}.} \bibinfo{year}{2016}\natexlab{}.
\newblock \showarticletitle{The experience of flow in artistic creation}.
\newblock \bibinfo{journal}{\emph{Flow experience: Empirical research and applications}} (\bibinfo{year}{2016}), \bibinfo{pages}{37--50}.
\newblock


\bibitem[Chen et~al\mbox{.}(2024)]%
        {FamilyWell-being}
\bibfield{author}{\bibinfo{person}{Yifei Chen}, \bibinfo{person}{Qinglin Mao}, \bibinfo{person}{Xinlie Huang}, {and} \bibinfo{person}{Ningning Xu}.} \bibinfo{year}{2024}\natexlab{}.
\newblock \showarticletitle{LanternOperAR: A Hybrid Cultural Gift for Quality Education and Family Well-being}. In \bibinfo{booktitle}{\emph{Extended Abstracts of the CHI Conference on Human Factors in Computing Systems}}. \bibinfo{pages}{1--8}.
\newblock


\bibitem[Christou et~al\mbox{.}(2006)]%
        {cavechristou2006versatile}
\bibfield{author}{\bibinfo{person}{Chris Christou}, \bibinfo{person}{Cameron Angus}, \bibinfo{person}{Celine Loscos}, \bibinfo{person}{Andrea Dettori}, {and} \bibinfo{person}{Maria Roussou}.} \bibinfo{year}{2006}\natexlab{}.
\newblock \showarticletitle{A versatile large-scale multimodal VR system for cultural heritage visualization}. In \bibinfo{booktitle}{\emph{Proceedings of the ACM symposium on Virtual reality software and technology}}. \bibinfo{pages}{133--140}.
\newblock


\bibitem[Colla{\c{c}}o et~al\mbox{.}(2021)]%
        {collacco2021immersion}
\bibfield{author}{\bibinfo{person}{Elen Colla{\c{c}}o}, \bibinfo{person}{Elisabeti Kira}, \bibinfo{person}{Lucas~H Sallaberry}, \bibinfo{person}{Anna~CM Queiroz}, \bibinfo{person}{Maria~AAM Machado}, \bibinfo{person}{Oswaldo Crivello~Jr}, {and} \bibinfo{person}{Romero Tori}.} \bibinfo{year}{2021}\natexlab{}.
\newblock \showarticletitle{Immersion and haptic feedback impacts on dental anesthesia technical skills virtual reality training}.
\newblock \bibinfo{journal}{\emph{Journal of Dental Education}} \bibinfo{volume}{85}, \bibinfo{number}{4} (\bibinfo{year}{2021}), \bibinfo{pages}{589--598}.
\newblock


\bibitem[Csikszentmihalyi et~al\mbox{.}(2005)]%
        {flow}
\bibfield{author}{\bibinfo{person}{Mihaly Csikszentmihalyi}, \bibinfo{person}{Sami Abuhamdeh}, {and} \bibinfo{person}{Jeanne Nakamura}.} \bibinfo{year}{2005}\natexlab{}.
\newblock \showarticletitle{Flow}.
\newblock \bibinfo{journal}{\emph{Handbook of competence and motivation}} (\bibinfo{year}{2005}), \bibinfo{pages}{598--608}.
\newblock


\bibitem[Dayarathna et~al\mbox{.}(2021)]%
        {dayarathna2021assessment}
\bibfield{author}{\bibinfo{person}{Vidanelage~L Dayarathna}, \bibinfo{person}{Sofia Karam}, \bibinfo{person}{Raed Jaradat}, \bibinfo{person}{Michael~A Hamilton}, \bibinfo{person}{Parker Jones}, \bibinfo{person}{Emily~S Wall}, \bibinfo{person}{Safae El~Amrani}, \bibinfo{person}{Niamat~Ullah Ibne~Hossain}, {and} \bibinfo{person}{Fatine Elakramine}.} \bibinfo{year}{2021}\natexlab{}.
\newblock \showarticletitle{An assessment of individuals’ systems thinking skills via immersive virtual reality complex system scenarios}.
\newblock \bibinfo{journal}{\emph{Systems}} \bibinfo{volume}{9}, \bibinfo{number}{2} (\bibinfo{year}{2021}), \bibinfo{pages}{40}.
\newblock


\bibitem[Dhawan et~al\mbox{.}(2016)]%
        {cricket}
\bibfield{author}{\bibinfo{person}{Aishwar Dhawan}, \bibinfo{person}{Alan Cummins}, \bibinfo{person}{Wayne Spratford}, \bibinfo{person}{Joost~C Dessing}, {and} \bibinfo{person}{Cathy Craig}.} \bibinfo{year}{2016}\natexlab{}.
\newblock \showarticletitle{Development of a novel immersive interactive virtual reality cricket simulator for cricket batting}. In \bibinfo{booktitle}{\emph{Proceedings of the 10th international Symposium on computer Science in sports (ISCSS)}}. Springer, \bibinfo{pages}{203--210}.
\newblock


\bibitem[Feinberg et~al\mbox{.}(2022)]%
        {zenvr_2022}
\bibfield{author}{\bibinfo{person}{Rachel~R. Feinberg}, \bibinfo{person}{Udaya Lakshmi}, \bibinfo{person}{Matthew~J. Golino}, {and} \bibinfo{person}{Rosa~I. Arriaga}.} \bibinfo{year}{2022}\natexlab{}.
\newblock \showarticletitle{{ZenVR}: Design Evaluation of a Virtual Reality Learning System for Meditation}. In \bibinfo{booktitle}{\emph{Proceedings of the 2022 {CHI} Conference on Human Factors in Computing Systems}} (New York, {NY}, {USA}) \emph{(\bibinfo{series}{{CHI} '22})}. \bibinfo{publisher}{Association for Computing Machinery}, \bibinfo{pages}{1--15}.
\newblock
\showISBNx{978-1-4503-9157-3}
\urldef\tempurl%
\url{https://doi.org/10.1145/3491102.3502035}
\showDOI{\tempurl}


\bibitem[Fernandez et~al\mbox{.}(2019)]%
        {amore2019deep}
\bibfield{author}{\bibinfo{person}{Judith~Amores Fernandez}, \bibinfo{person}{Anna Fust\'{e}}, \bibinfo{person}{Robert Richer}, {and} \bibinfo{person}{Pattie Maes}.} \bibinfo{year}{2019}\natexlab{}.
\newblock \showarticletitle{Deep reality: an underwater VR experience to promote relaxation by unconscious HR, EDA, and brain activity biofeedback}. In \bibinfo{booktitle}{\emph{ACM SIGGRAPH 2019 Virtual, Augmented, and Mixed Reality}} (Los Angeles, California) \emph{(\bibinfo{series}{SIGGRAPH '19})}. \bibinfo{publisher}{Association for Computing Machinery}, \bibinfo{address}{New York, NY, USA}, Article \bibinfo{articleno}{17}, \bibinfo{numpages}{1}~pages.
\newblock
\showISBNx{9781450363204}


\bibitem[Francis et~al\mbox{.}(2012)]%
        {bonetechnical}
\bibfield{author}{\bibinfo{person}{Howard~W Francis}, \bibinfo{person}{Mohammad~U Malik}, \bibinfo{person}{David~A Diaz Voss~Varela}, \bibinfo{person}{Maxwell~A Barffour}, \bibinfo{person}{Wade~W Chien}, \bibinfo{person}{John~P Carey}, \bibinfo{person}{John~K Niparko}, {and} \bibinfo{person}{Nasir~I Bhatti}.} \bibinfo{year}{2012}\natexlab{}.
\newblock \showarticletitle{Technical skills improve after practice on virtual-reality temporal bone simulator}.
\newblock \bibinfo{journal}{\emph{The Laryngoscope}} \bibinfo{volume}{122}, \bibinfo{number}{6} (\bibinfo{year}{2012}), \bibinfo{pages}{1385--1391}.
\newblock


\bibitem[Gaertner et~al\mbox{.}(2023)]%
        {gaertner2023relaxing}
\bibfield{author}{\bibinfo{person}{Raphaela~J Gaertner}, \bibinfo{person}{Katharina~E Kossmann}, \bibinfo{person}{Annika~BE Benz}, \bibinfo{person}{Ulrike~U Bentele}, \bibinfo{person}{Maria Meier}, \bibinfo{person}{Bernadette~F Denk}, \bibinfo{person}{Elea~SC Klink}, \bibinfo{person}{Stephanie~J Dimitroff}, {and} \bibinfo{person}{Jens~C Pruessner}.} \bibinfo{year}{2023}\natexlab{}.
\newblock \showarticletitle{Relaxing effects of virtual environments on the autonomic nervous system indicated by heart rate variability: A systematic review}.
\newblock \bibinfo{journal}{\emph{Journal of Environmental Psychology}}  \bibinfo{volume}{88} (\bibinfo{year}{2023}), \bibinfo{pages}{102035}.
\newblock


\bibitem[Gao et~al\mbox{.}(2023)]%
        {bamboo}
\bibfield{author}{\bibinfo{person}{Peizhong Gao}, \bibinfo{person}{Tanhao Gao}, \bibinfo{person}{Yanbin Yang}, \bibinfo{person}{Zhenyuan Liu}, \bibinfo{person}{Jianyu Shi}, {and} \bibinfo{person}{Jin Li}.} \bibinfo{year}{2023}\natexlab{}.
\newblock \showarticletitle{Bamboo Agents: Exploring the Potentiality of Digital Craft by Decoding and Recoding Process}. In \bibinfo{booktitle}{\emph{Proceedings of the Seventeenth International Conference on Tangible, Embedded, and Embodied Interaction}}. \bibinfo{pages}{1--13}.
\newblock


\bibitem[Harmon-Jones et~al\mbox{.}(2016)]%
        {discreteharmon2016}
\bibfield{author}{\bibinfo{person}{Cindy Harmon-Jones}, \bibinfo{person}{Brock Bastian}, {and} \bibinfo{person}{Eddie Harmon-Jones}.} \bibinfo{year}{2016}\natexlab{}.
\newblock \showarticletitle{The discrete emotions questionnaire: A new tool for measuring state self-reported emotions}.
\newblock \bibinfo{journal}{\emph{PloS one}} \bibinfo{volume}{11}, \bibinfo{number}{8} (\bibinfo{year}{2016}), \bibinfo{pages}{e0159915}.
\newblock


\bibitem[Hart(1988)]%
        {hart1988development}
\bibfield{author}{\bibinfo{person}{S Hart}.} \bibinfo{year}{1988}\natexlab{}.
\newblock \showarticletitle{Development of NASA-TLX (Task Load Index): Results of empirical and theoretical research}.
\newblock \bibinfo{journal}{\emph{Human mental workload/Elsevier}} (\bibinfo{year}{1988}).
\newblock


\bibitem[Harvey et~al\mbox{.}(2021)]%
        {motorskill1}
\bibfield{author}{\bibinfo{person}{Carlo Harvey}, \bibinfo{person}{Elmedin Selmanovi{\'c}}, \bibinfo{person}{Jake O'Connor}, {and} \bibinfo{person}{Malek Chahin}.} \bibinfo{year}{2021}\natexlab{}.
\newblock \showarticletitle{A comparison between expert and beginner learning for motor skill development in a virtual reality serious game}.
\newblock \bibinfo{journal}{\emph{The Visual Computer}} \bibinfo{volume}{37}, \bibinfo{number}{1} (\bibinfo{year}{2021}), \bibinfo{pages}{3--17}.
\newblock


\bibitem[Hongmei(2023)]%
        {ideological}
\bibfield{author}{\bibinfo{person}{Jing Hongmei}.} \bibinfo{year}{2023}\natexlab{}.
\newblock \showarticletitle{The ideological origins and aesthetic construction of yijing (artistic conception)}.
\newblock \bibinfo{journal}{\emph{International Communication of Chinese Culture}} \bibinfo{volume}{10}, \bibinfo{number}{2} (\bibinfo{year}{2023}), \bibinfo{pages}{151--169}.
\newblock


\bibitem[Hou et~al\mbox{.}(2022)]%
        {digitizing_2022}
\bibfield{author}{\bibinfo{person}{Yumeng Hou}, \bibinfo{person}{Sarah Kenderdine}, \bibinfo{person}{Davide Picca}, \bibinfo{person}{Mattia Egloff}, {and} \bibinfo{person}{Alessandro Adamou}.} \bibinfo{year}{2022}\natexlab{}.
\newblock \showarticletitle{Digitizing intangible cultural heritage embodied: State of the art}.
\newblock \bibinfo{journal}{\emph{Journal on Computing and Cultural Heritage (JOCCH)}} \bibinfo{volume}{15}, \bibinfo{number}{3} (\bibinfo{year}{2022}), \bibinfo{pages}{1--20}.
\newblock
\urldef\tempurl%
\url{https://doi.org/10.1145/3494837}
\showDOI{\tempurl}


\bibitem[Huang and Lee(2022)]%
        {huang2022factors}
\bibfield{author}{\bibinfo{person}{Hsinfu Huang} {and} \bibinfo{person}{Chang-Franw Lee}.} \bibinfo{year}{2022}\natexlab{}.
\newblock \showarticletitle{Factors affecting usability of 3D model learning in a virtual reality environment}.
\newblock \bibinfo{journal}{\emph{Interactive Learning Environments}} \bibinfo{volume}{30}, \bibinfo{number}{5} (\bibinfo{year}{2022}), \bibinfo{pages}{848--861}.
\newblock


\bibitem[Jacobs and Zoran(2015)]%
        {Hunter-Gatherer}
\bibfield{author}{\bibinfo{person}{Jennifer Jacobs} {and} \bibinfo{person}{Amit Zoran}.} \bibinfo{year}{2015}\natexlab{}.
\newblock \showarticletitle{Hybrid Practice in the Kalahari: Design Collaboration through Digital Tools and Hunter-Gatherer Craft.}. In \bibinfo{booktitle}{\emph{CHI}}, Vol.~\bibinfo{volume}{15}. \bibinfo{pages}{619--628}.
\newblock


\bibitem[Ji et~al\mbox{.}(2019)]%
        {ShanghaiStyleLacquerware}
\bibfield{author}{\bibinfo{person}{Jingyi Ji}, \bibinfo{person}{Jianxin Cheng}, {and} \bibinfo{person}{Rongrong Fu}.} \bibinfo{year}{2019}\natexlab{}.
\newblock \showarticletitle{Research on Design Process of Small Intangible Cultural Heritage Art Gallery Based on VRP-MUSEUM Technology—Taking the Art Gallery of Shanghai Style Lacquerware as an Example}. In \bibinfo{booktitle}{\emph{HCI International 2019-Posters: 21st International Conference, HCII 2019, Orlando, FL, USA, July 26--31, 2019, Proceedings, Part III 21}}. Springer, \bibinfo{pages}{566--574}.
\newblock


\bibitem[Johnson-Glenberg(2018)]%
        {Embodied}
\bibfield{author}{\bibinfo{person}{Mina~C Johnson-Glenberg}.} \bibinfo{year}{2018}\natexlab{}.
\newblock \showarticletitle{Immersive VR and education: Embodied design principles that include gesture and hand controls}.
\newblock \bibinfo{journal}{\emph{Frontiers in Robotics and AI}}  \bibinfo{volume}{5} (\bibinfo{year}{2018}), \bibinfo{pages}{375272}.
\newblock


\bibitem[Jurgaitis et~al\mbox{.}(2008)]%
        {comparison2-dimensional}
\bibfield{author}{\bibinfo{person}{Jonas Jurgaitis}, \bibinfo{person}{Marius Pa{\v{s}}konis}, \bibinfo{person}{Jonas Pivori{\=u}nas}, \bibinfo{person}{Ieva Martinaityt{\.e}}, \bibinfo{person}{Agnius Ju{\v{s}}ka}, \bibinfo{person}{R{\=u}ta Jurgaitien{\.e}}, \bibinfo{person}{Art{\=u}ras Samuilis}, \bibinfo{person}{Ivo Volf}, \bibinfo{person}{Maks Sch{\"o}binger}, \bibinfo{person}{Peter Schemmer}, {et~al\mbox{.}}} \bibinfo{year}{2008}\natexlab{}.
\newblock \showarticletitle{The comparison of 2-dimensional with 3-dimensional hepatic visualization in the clinical hepatic anatomy education}.
\newblock \bibinfo{journal}{\emph{Medicina}} \bibinfo{volume}{44}, \bibinfo{number}{6} (\bibinfo{year}{2008}), \bibinfo{pages}{428}.
\newblock


\bibitem[Khasky and Smith(1999)]%
        {1999stress}
\bibfield{author}{\bibinfo{person}{Amy~D Khasky} {and} \bibinfo{person}{Jonathan~C Smith}.} \bibinfo{year}{1999}\natexlab{}.
\newblock \showarticletitle{Stress, relaxation states, and creativity}.
\newblock \bibinfo{journal}{\emph{Perceptual and motor skills}} \bibinfo{volume}{88}, \bibinfo{number}{2} (\bibinfo{year}{1999}), \bibinfo{pages}{409--416}.
\newblock


\bibitem[Kohut(2009)]%
        {kohut2009restoration}
\bibfield{author}{\bibinfo{person}{Heinz Kohut}.} \bibinfo{year}{2009}\natexlab{}.
\newblock \bibinfo{booktitle}{\emph{The restoration of the self}}.
\newblock \bibinfo{publisher}{University of Chicago Press}.
\newblock


\bibitem[Kolb(2014)]%
        {kolb2014experiential}
\bibfield{author}{\bibinfo{person}{David~A Kolb}.} \bibinfo{year}{2014}\natexlab{}.
\newblock \bibinfo{booktitle}{\emph{Experiential learning: Experience as the source of learning and development}}.
\newblock \bibinfo{publisher}{FT press}.
\newblock


\bibitem[Kurin et~al\mbox{.}(2007)]%
        {kurin2007safeguarding}
\bibfield{author}{\bibinfo{person}{Richard Kurin} {et~al\mbox{.}}} \bibinfo{year}{2007}\natexlab{}.
\newblock \showarticletitle{Safeguarding intangible cultural heritage: Key factors in implementing the 2003 Convention}.
\newblock \bibinfo{journal}{\emph{International journal of intangible heritage}} \bibinfo{volume}{2}, \bibinfo{number}{8} (\bibinfo{year}{2007}), \bibinfo{pages}{9--20}.
\newblock


\bibitem[Lehrer(2007)]%
        {biofeedback}
\bibfield{author}{\bibinfo{person}{Paul~M Lehrer}.} \bibinfo{year}{2007}\natexlab{}.
\newblock \showarticletitle{Biofeedback training to increase heart rate variability}.
\newblock \bibinfo{journal}{\emph{Principles and practice of stress management}}  \bibinfo{volume}{3} (\bibinfo{year}{2007}), \bibinfo{pages}{227--248}.
\newblock


\bibitem[Lenzerini(2011)]%
        {livingintangible_2011}
\bibfield{author}{\bibinfo{person}{Federico Lenzerini}.} \bibinfo{year}{2011}\natexlab{}.
\newblock \showarticletitle{Intangible Cultural Heritage: The Living Culture of Peoples}.
\newblock \bibinfo{journal}{\emph{European Journal of International Law}} \bibinfo{volume}{22}, \bibinfo{number}{1} (\bibinfo{year}{2011}), \bibinfo{pages}{101--120}.
\newblock
\showISSN{0938-5428}
\urldef\tempurl%
\url{https://doi.org/10.1093/ejil/chr006}
\showDOI{\tempurl}


\bibitem[Li(2002)]%
        {li2002chinese}
\bibfield{author}{\bibinfo{person}{Hui-Lin Li}.} \bibinfo{year}{2002}\natexlab{}.
\newblock \bibinfo{booktitle}{\emph{Chinese flower arrangement}}.
\newblock \bibinfo{publisher}{Courier Corporation}.
\newblock


\bibitem[Li et~al\mbox{.}(2023)]%
        {li2023diantea}
\bibfield{author}{\bibinfo{person}{Jiajia Li}, \bibinfo{person}{Zixia Zheng}, \bibinfo{person}{Yaqing Chai}, \bibinfo{person}{Shizhen Su}, \bibinfo{person}{Xiemin Wei}, \bibinfo{person}{Hongning Shi}, {and} \bibinfo{person}{Xiangyang Xin}.} \bibinfo{year}{2023}\natexlab{}.
\newblock \showarticletitle{DianTea: designing and evaluating an immersive virtual reality game to enhance youth tea culture learning}. In \bibinfo{booktitle}{\emph{Proceedings of the 25th International Conference on Mobile Human-Computer Interaction}}. \bibinfo{pages}{1--8}.
\newblock


\bibitem[Li et~al\mbox{.}(2022)]%
        {li2022electrodermal}
\bibfield{author}{\bibinfo{person}{Shanshi Li}, \bibinfo{person}{Billy Sung}, \bibinfo{person}{Yuxia Lin}, {and} \bibinfo{person}{Ondrej Mitas}.} \bibinfo{year}{2022}\natexlab{}.
\newblock \showarticletitle{Electrodermal activity measure: A methodological review}.
\newblock \bibinfo{journal}{\emph{Annals of Tourism Research}}  \bibinfo{volume}{96} (\bibinfo{year}{2022}), \bibinfo{pages}{103460}.
\newblock


\bibitem[Liu et~al\mbox{.}(2023)]%
        {hairymonkey}
\bibfield{author}{\bibinfo{person}{Guanhong Liu}, \bibinfo{person}{Xianghua Ding}, \bibinfo{person}{Jinghe Cai}, \bibinfo{person}{Weiyun Wang}, \bibinfo{person}{Xinyue Wang}, \bibinfo{person}{Yuting Diao}, \bibinfo{person}{Jin Chen}, \bibinfo{person}{Tianyu Yu}, \bibinfo{person}{Haiqing Xu}, {and} \bibinfo{person}{Haipeng Mi}.} \bibinfo{year}{2023}\natexlab{}.
\newblock \showarticletitle{Digital making for inheritance and enlivening intangible cultural heritage: A case of hairy monkey handicrafts}. In \bibinfo{booktitle}{\emph{Proceedings of the 2023 CHI conference on human factors in computing systems}}. \bibinfo{pages}{1--17}.
\newblock


\bibitem[Liu et~al\mbox{.}(2024)]%
        {HybridCraft}
\bibfield{author}{\bibinfo{person}{Guanhong Liu}, \bibinfo{person}{Qingyuan Shi}, \bibinfo{person}{Yuan Yao}, \bibinfo{person}{Yuan-Ling Feng}, \bibinfo{person}{Tianyu Yu}, \bibinfo{person}{Beituo Liu}, \bibinfo{person}{Zhijun Ma}, \bibinfo{person}{Li Huang}, {and} \bibinfo{person}{Yuting Diao}.} \bibinfo{year}{2024}\natexlab{}.
\newblock \showarticletitle{Learning from Hybrid Craft: Investigating and Reflecting on Innovating and Enlivening Traditional Craft through Literature Review}. In \bibinfo{booktitle}{\emph{Proceedings of the CHI Conference on Human Factors in Computing Systems}}. \bibinfo{pages}{1--19}.
\newblock


\bibitem[Liu et~al\mbox{.}(2022)]%
        {huaer}
\bibfield{author}{\bibinfo{person}{Zixiao Liu}, \bibinfo{person}{Shuo Yan}, \bibinfo{person}{Yu Lu}, {and} \bibinfo{person}{Yuetong Zhao}.} \bibinfo{year}{2022}\natexlab{}.
\newblock \showarticletitle{Generating Embodied Storytelling and Interactive Experience of China Intangible Cultural Heritage “Hua'er” in Virtual Reality}. In \bibinfo{booktitle}{\emph{CHI Conference on Human Factors in Computing Systems Extended Abstracts}}. \bibinfo{pages}{1--7}.
\newblock


\bibitem[Lovasz-Bukvova et~al\mbox{.}(2021)]%
        {lovasz2021usability}
\bibfield{author}{\bibinfo{person}{Helena Lovasz-Bukvova}, \bibinfo{person}{Marvin H{\"o}lzl}, \bibinfo{person}{Gerhard Kormann-Hainzl}, \bibinfo{person}{Thomas Moser}, \bibinfo{person}{Tanja Zigart}, {and} \bibinfo{person}{Sebastian Schlund}.} \bibinfo{year}{2021}\natexlab{}.
\newblock \showarticletitle{Usability and Task Load of Applications in Augmented and Virtual Reality: How Applicable are the Technologies in Corporate Settings?}. In \bibinfo{booktitle}{\emph{Systems, Software and Services Process Improvement: 28th European Conference, EuroSPI 2021, Krems, Austria, September 1--3, 2021, Proceedings 28}}. Springer, \bibinfo{pages}{708--718}.
\newblock


\bibitem[Nassar et~al\mbox{.}(2021)]%
        {simulationmodality}
\bibfield{author}{\bibinfo{person}{Aussama~K Nassar}, \bibinfo{person}{Farris Al-Manaseer}, \bibinfo{person}{Lisa~M Knowlton}, {and} \bibinfo{person}{Faiz Tuma}.} \bibinfo{year}{2021}\natexlab{}.
\newblock \showarticletitle{Virtual reality (VR) as a simulation modality for technical skills acquisition}.
\newblock \bibinfo{journal}{\emph{Annals of Medicine and Surgery}}  \bibinfo{volume}{71} (\bibinfo{year}{2021}).
\newblock


\bibitem[Ning(2023)]%
        {ning2023analysis}
\bibfield{author}{\bibinfo{person}{Hui Ning}.} \bibinfo{year}{2023}\natexlab{}.
\newblock \showarticletitle{Analysis of the value of folk music intangible cultural heritage on the regulation of mental health}.
\newblock \bibinfo{journal}{\emph{Frontiers in Psychiatry}}  \bibinfo{volume}{14} (\bibinfo{year}{2023}), \bibinfo{pages}{1067753}.
\newblock


\bibitem[Ntagiantas et~al\mbox{.}(2021)]%
        {book}
\bibfield{author}{\bibinfo{person}{Antonios Ntagiantas}, \bibinfo{person}{Dimitris Manousos}, \bibinfo{person}{Markos Konstantakis}, \bibinfo{person}{John Aliprantis}, {and} \bibinfo{person}{George Caridakis}.} \bibinfo{year}{2021}\natexlab{}.
\newblock \showarticletitle{Augmented Reality children's book for intangible cultural heritage through participatory content creation and promotion. Case study: the pastoral life of Psiloritis as a UNESCO World Geopark}. In \bibinfo{booktitle}{\emph{2021 16th International Workshop on Semantic and Social Media Adaptation \& Personalization (SMAP)}}. \bibinfo{pages}{1--4}.
\newblock
\urldef\tempurl%
\url{https://doi.org/10.1109/SMAP53521.2021.9610762}
\showDOI{\tempurl}


\bibitem[Oktay(2012)]%
        {oktay2012grounded}
\bibfield{author}{\bibinfo{person}{Julianne~S Oktay}.} \bibinfo{year}{2012}\natexlab{}.
\newblock \bibinfo{booktitle}{\emph{Grounded theory}}.
\newblock \bibinfo{publisher}{Oxford University Press}.
\newblock


\bibitem[Papachristopoulos et~al\mbox{.}(2023)]%
        {papachristopoulos2023positive}
\bibfield{author}{\bibinfo{person}{Konstantinos Papachristopoulos}, \bibinfo{person}{Marc-Antoine Gradito~Dubord}, \bibinfo{person}{Florence Jauvin}, \bibinfo{person}{Jacques Forest}, {and} \bibinfo{person}{Patrick Coulombe}.} \bibinfo{year}{2023}\natexlab{}.
\newblock \showarticletitle{Positive Impact, Creativity, and Innovative Behavior at Work: The Mediating Role of Basic Needs Satisfaction}.
\newblock \bibinfo{journal}{\emph{Behavioral Sciences}} \bibinfo{volume}{13}, \bibinfo{number}{12} (\bibinfo{year}{2023}), \bibinfo{pages}{984}.
\newblock


\bibitem[Partarakis et~al\mbox{.}(2021)]%
        {role-play}
\bibfield{author}{\bibinfo{person}{Nikolaos Partarakis}, \bibinfo{person}{Nikolaos Patsiouras}, \bibinfo{person}{Thodoris Evdemon}, \bibinfo{person}{Paraskevi Doulgeraki}, \bibinfo{person}{Effie Karuzaki}, \bibinfo{person}{Evropi Stefanidi}, \bibinfo{person}{Stavroula Ntoa}, \bibinfo{person}{Carlo Meghini}, \bibinfo{person}{Danai Kaplanidi}, \bibinfo{person}{Maria Fasoula}, {et~al\mbox{.}}} \bibinfo{year}{2021}\natexlab{}.
\newblock \showarticletitle{Enhancing the educational value of tangible and intangible dimensions of traditional crafts through role-play gaming}. In \bibinfo{booktitle}{\emph{Interactivity and Game Creation: 9th EAI International Conference, ArtsIT 2020, Aalborg, Denmark, December 10--11, 2020, Proceedings 9}}. Springer, \bibinfo{pages}{243--254}.
\newblock


\bibitem[Partarakis et~al\mbox{.}(2020)]%
        {traditionalcrafts}
\bibfield{author}{\bibinfo{person}{Nikolaos Partarakis}, \bibinfo{person}{Xenophon Zabulis}, \bibinfo{person}{Antonis Chatziantoniou}, \bibinfo{person}{Nikolaos Patsiouras}, {and} \bibinfo{person}{Ilia Adami}.} \bibinfo{year}{2020}\natexlab{}.
\newblock \showarticletitle{An approach to the creation and presentation of reference gesture datasets, for the preservation of traditional crafts}.
\newblock \bibinfo{journal}{\emph{Applied Sciences}} \bibinfo{volume}{10}, \bibinfo{number}{20} (\bibinfo{year}{2020}), \bibinfo{pages}{7325}.
\newblock


\bibitem[Petronela(2016)]%
        {importance_2016}
\bibfield{author}{\bibinfo{person}{Tudorache Petronela}.} \bibinfo{year}{2016}\natexlab{}.
\newblock \showarticletitle{The Importance of the Intangible Cultural Heritage in the Economy}.
\newblock \bibinfo{journal}{\emph{Procedia Economics and Finance}}  \bibinfo{volume}{39} (\bibinfo{year}{2016}), \bibinfo{pages}{731--736}.
\newblock
\showISSN{2212-5671}
\urldef\tempurl%
\url{https://doi.org/10.1016/S2212-5671(16)30271-4}
\showDOI{\tempurl}


\bibitem[Rallis et~al\mbox{.}(2020)]%
        {danceanalysis}
\bibfield{author}{\bibinfo{person}{Ioannis Rallis}, \bibinfo{person}{Athanasios Voulodimos}, \bibinfo{person}{Nikolaos Bakalos}, \bibinfo{person}{Eftychios Protopapadakis}, \bibinfo{person}{Nikolaos Doulamis}, {and} \bibinfo{person}{Anastasios Doulamis}.} \bibinfo{year}{2020}\natexlab{}.
\newblock \showarticletitle{Machine learning for intangible cultural heritage: a review of techniques on dance analysis}.
\newblock \bibinfo{journal}{\emph{Visual Computing for Cultural Heritage}} (\bibinfo{year}{2020}), \bibinfo{pages}{103--119}.
\newblock


\bibitem[Ruvimova et~al\mbox{.}(2020)]%
        {Transport}
\bibfield{author}{\bibinfo{person}{Anastasia Ruvimova}, \bibinfo{person}{Junhyeok Kim}, \bibinfo{person}{Thomas Fritz}, \bibinfo{person}{Mark Hancock}, {and} \bibinfo{person}{David~C. Shepherd}.} \bibinfo{year}{2020}\natexlab{}.
\newblock \showarticletitle{"Transport Me Away": Fostering Flow in Open Offices through Virtual Reality}. In \bibinfo{booktitle}{\emph{Proceedings of the 2020 CHI Conference on Human Factors in Computing Systems}} (Honolulu, HI, USA) \emph{(\bibinfo{series}{CHI '20})}. \bibinfo{publisher}{Association for Computing Machinery}, \bibinfo{address}{New York, NY, USA}, \bibinfo{pages}{1–14}.
\newblock
\showISBNx{9781450367080}
\urldef\tempurl%
\url{https://doi.org/10.1145/3313831.3376724}
\showDOI{\tempurl}


\bibitem[Sabie et~al\mbox{.}(2023)]%
        {interculturalheritage}
\bibfield{author}{\bibinfo{person}{Dina Sabie}, \bibinfo{person}{Hala Sheta}, \bibinfo{person}{Hasan~Shahid Ferdous}, \bibinfo{person}{Vannie Kopalakrishnan}, {and} \bibinfo{person}{Syed~Ishtiaque Ahmed}.} \bibinfo{year}{2023}\natexlab{}.
\newblock \showarticletitle{Be our guest: intercultural heritage exchange through augmented reality (AR)}. In \bibinfo{booktitle}{\emph{Proceedings of the 2023 CHI conference on human factors in computing systems}}. \bibinfo{pages}{1--15}.
\newblock


\bibitem[Samaniego et~al\mbox{.}(2024)]%
        {samaniego2024creative}
\bibfield{author}{\bibinfo{person}{Mariela Samaniego}, \bibinfo{person}{Nancy Usca}, \bibinfo{person}{Jos{\'e} Salguero}, {and} \bibinfo{person}{William Quevedo}.} \bibinfo{year}{2024}\natexlab{}.
\newblock \showarticletitle{Creative Thinking in Art and Design Education: A Systematic Review}.
\newblock \bibinfo{journal}{\emph{Education Sciences}} \bibinfo{volume}{14}, \bibinfo{number}{2} (\bibinfo{year}{2024}), \bibinfo{pages}{192}.
\newblock


\bibitem[Sayer(2015)]%
        {sayer2015can}
\bibfield{author}{\bibinfo{person}{Faye Sayer}.} \bibinfo{year}{2015}\natexlab{}.
\newblock \showarticletitle{Can digging make you happy? Archaeological excavations, happiness and heritage}.
\newblock \bibinfo{journal}{\emph{Arts \& Health}} \bibinfo{volume}{7}, \bibinfo{number}{3} (\bibinfo{year}{2015}), \bibinfo{pages}{247--260}.
\newblock


\bibitem[Seabrook et~al\mbox{.}(2020)]%
        {mindfulnesspractice}
\bibfield{author}{\bibinfo{person}{Elizabeth Seabrook}, \bibinfo{person}{Ryan Kelly}, \bibinfo{person}{Fiona Foley}, \bibinfo{person}{Stephen Theiler}, \bibinfo{person}{Neil Thomas}, \bibinfo{person}{Greg Wadley}, {and} \bibinfo{person}{Maja Nedeljkovic}.} \bibinfo{year}{2020}\natexlab{}.
\newblock \showarticletitle{Understanding how virtual reality can support mindfulness practice: mixed methods study}.
\newblock \bibinfo{journal}{\emph{Journal of medical Internet research}} \bibinfo{volume}{22}, \bibinfo{number}{3} (\bibinfo{year}{2020}), \bibinfo{pages}{e16106}.
\newblock


\bibitem[Selmanovic et~al\mbox{.}(2018)]%
        {videostory2018vr}
\bibfield{author}{\bibinfo{person}{Elmedin Selmanovic}, \bibinfo{person}{Selma Rizvic}, \bibinfo{person}{Carlo Harvey}, \bibinfo{person}{Dusanka Boskovic}, \bibinfo{person}{Vedad Hulusic}, \bibinfo{person}{Malek Chahin}, {and} \bibinfo{person}{Sanda Sljivo}.} \bibinfo{year}{2018}\natexlab{}.
\newblock \showarticletitle{{VR Video Storytelling for Intangible Cultural Heritage Preservation}}. In \bibinfo{booktitle}{\emph{Eurographics Workshop on Graphics and Cultural Heritage}}, \bibfield{editor}{\bibinfo{person}{Robert Sablatnig} {and} \bibinfo{person}{Michael Wimmer}} (Eds.). \bibinfo{publisher}{The Eurographics Association}.
\newblock
\showISBNx{978-3-03868-057-4}
\showISSN{2312-6124}
\urldef\tempurl%
\url{https://doi.org/10.2312/gch.20181341}
\showDOI{\tempurl}


\bibitem[Selmanovi\'{c} et~al\mbox{.}(2020)]%
        {Accessibility}
\bibfield{author}{\bibinfo{person}{Elmedin Selmanovi\'{c}}, \bibinfo{person}{Selma Rizvic}, \bibinfo{person}{Carlo Harvey}, \bibinfo{person}{Dusanka Boskovic}, \bibinfo{person}{Vedad Hulusic}, \bibinfo{person}{Malek Chahin}, {and} \bibinfo{person}{Sanda Sljivo}.} \bibinfo{year}{2020}\natexlab{}.
\newblock \showarticletitle{Improving Accessibility to Intangible Cultural Heritage Preservation Using Virtual Reality}.
\newblock \bibinfo{journal}{\emph{J. Comput. Cult. Herit.}} \bibinfo{volume}{13}, \bibinfo{number}{2}, Article \bibinfo{articleno}{13} (\bibinfo{date}{May} \bibinfo{year}{2020}), \bibinfo{numpages}{19}~pages.
\newblock
\showISSN{1556-4673}
\urldef\tempurl%
\url{https://doi.org/10.1145/3377143}
\showDOI{\tempurl}


\bibitem[Shin and Woo(2023)]%
        {Storytellingspace}
\bibfield{author}{\bibinfo{person}{Jae-Eun Shin} {and} \bibinfo{person}{Woontack Woo}.} \bibinfo{year}{2023}\natexlab{}.
\newblock \showarticletitle{How Space is Told: Linking Trajectory, Narrative, and Intent in Augmented Reality Storytelling for Cultural Heritage Sites}. In \bibinfo{booktitle}{\emph{Proceedings of the 2023 CHI Conference on Human Factors in Computing Systems}}. \bibinfo{pages}{1--14}.
\newblock


\bibitem[Skovfoged et~al\mbox{.}(2018)]%
        {2018tales}
\bibfield{author}{\bibinfo{person}{Milo~Marsfeldt Skovfoged}, \bibinfo{person}{Martin Viktor}, \bibinfo{person}{Miroslav~Kalinov Sokolov}, \bibinfo{person}{Anders Hansen}, \bibinfo{person}{Helene~H{\o}gh Nielsen}, {and} \bibinfo{person}{Kasper Rodil}.} \bibinfo{year}{2018}\natexlab{}.
\newblock \showarticletitle{The tales of the Tokoloshe: Safeguarding intangible cultural heritage using virtual reality}. In \bibinfo{booktitle}{\emph{Proceedings of the Second African Conference for Human Computer Interaction: Thriving Communities}}. \bibinfo{pages}{1--4}.
\newblock


\bibitem[Smith et~al\mbox{.}(2000)]%
        {ABCsmith2000abc}
\bibfield{author}{\bibinfo{person}{Jonathan~C Smith}, \bibinfo{person}{Amy~B Wedell}, \bibinfo{person}{Camille~J Kolotylo}, \bibinfo{person}{Jacquie~E Lewis}, \bibinfo{person}{Kristie~Y Byers}, {and} \bibinfo{person}{Christine~M Segin}.} \bibinfo{year}{2000}\natexlab{}.
\newblock \showarticletitle{ABC relaxation theory and the factor structure of relaxation states, recalled relaxation activities, dispositions, and motivations}.
\newblock \bibinfo{journal}{\emph{Psychological Reports}} \bibinfo{volume}{86}, \bibinfo{number}{3\_suppl} (\bibinfo{year}{2000}), \bibinfo{pages}{1201--1208}.
\newblock


\bibitem[Smuka(2016)]%
        {locationsmuka2016intangible}
\bibfield{author}{\bibinfo{person}{Ingrida Smuka}.} \bibinfo{year}{2016}\natexlab{}.
\newblock \showarticletitle{Intangible cultural heritage in promotion of development of location}. In \bibinfo{booktitle}{\emph{Economic Science for Rural Development Conference Proceedings}}. \bibinfo{pages}{7}.
\newblock


\bibitem[Som et~al\mbox{.}(2023)]%
        {som2023virtual}
\bibfield{author}{\bibinfo{person}{Sumana Som}, \bibinfo{person}{Deepak~John Mathew}, {and} \bibinfo{person}{Kim Vincs}.} \bibinfo{year}{2023}\natexlab{}.
\newblock \showarticletitle{Virtual reality for creativity practice and art and design education: a literature review}. In \bibinfo{booktitle}{\emph{International Conference on Research into Design}}. Springer, \bibinfo{pages}{1011--1022}.
\newblock


\bibitem[Spielberger et~al\mbox{.}(1971)]%
        {spielberger1971state}
\bibfield{author}{\bibinfo{person}{Charles~D Spielberger}, \bibinfo{person}{Fernando Gonzalez-Reigosa}, \bibinfo{person}{Angel Martinez-Urrutia}, \bibinfo{person}{Luiz~FS Natalicio}, {and} \bibinfo{person}{Diana~S Natalicio}.} \bibinfo{year}{1971}\natexlab{}.
\newblock \showarticletitle{The state-trait anxiety inventory}.
\newblock \bibinfo{journal}{\emph{Revista Interamericana de Psicologia/Interamerican journal of psychology}} \bibinfo{volume}{5}, \bibinfo{number}{3 \& 4} (\bibinfo{year}{1971}).
\newblock


\bibitem[Steghaus and Poth(2022)]%
        {assessingsteghaus2022}
\bibfield{author}{\bibinfo{person}{Sarah Steghaus} {and} \bibinfo{person}{Christian~H Poth}.} \bibinfo{year}{2022}\natexlab{}.
\newblock \showarticletitle{Assessing momentary relaxation using the Relaxation State Questionnaire (RSQ)}.
\newblock \bibinfo{journal}{\emph{Scientific Reports}} \bibinfo{volume}{12}, \bibinfo{number}{1} (\bibinfo{year}{2022}), \bibinfo{pages}{16341}.
\newblock


\bibitem[Sun et~al\mbox{.}(2021)]%
        {Apprenticessun_research_2021}
\bibfield{author}{\bibinfo{person}{Changqing Sun}, \bibinfo{person}{Hong Chen}, {and} \bibinfo{person}{Ruihua Liao}.} \bibinfo{year}{2021}\natexlab{}.
\newblock \showarticletitle{Research on Incentive Mechanism and Strategy Choice for Passing on Intangible Cultural Heritage from Masters to Apprentices}.
\newblock \bibinfo{journal}{\emph{Sustainability}} \bibinfo{volume}{13}, \bibinfo{number}{9} (\bibinfo{year}{2021}), \bibinfo{pages}{5245}.
\newblock
\showISSN{2071-1050}
\urldef\tempurl%
\url{https://doi.org/10.3390/su13095245}
\showDOI{\tempurl}
\newblock
\shownote{Number: 9 Publisher: Multidisciplinary Digital Publishing Institute}.


\bibitem[Sun et~al\mbox{.}(2024)]%
        {restoringsun2024}
\bibfield{author}{\bibinfo{person}{Tongxin Sun}, \bibinfo{person}{Tongtong Jin}, \bibinfo{person}{Yuru Huang}, \bibinfo{person}{Meng Li}, \bibinfo{person}{Yun Wang}, \bibinfo{person}{Zhe Jia}, {and} \bibinfo{person}{Xinyi Fu}.} \bibinfo{year}{2024}\natexlab{}.
\newblock \showarticletitle{Restoring dunhuang murals: crafting cultural heritage preservation knowledge into immersive virtual reality experience design}.
\newblock \bibinfo{journal}{\emph{International Journal of Human--Computer Interaction}} \bibinfo{volume}{40}, \bibinfo{number}{8} (\bibinfo{year}{2024}), \bibinfo{pages}{2019--2040}.
\newblock


\bibitem[UNESCO(2011)]%
        {WhatisIntangible}
\bibfield{author}{\bibinfo{person}{UNESCO}.} \bibinfo{year}{2011}\natexlab{}.
\newblock \bibinfo{booktitle}{\emph{{UNESCO} - What is Intangible Cultural Heritage?}}
\newblock UNESCO.
\newblock
\urldef\tempurl%
\url{https://ich.unesco.org/en/what-is-intangible-heritage-00003}
\showURL{%
\tempurl}


\bibitem[{UNESCO Intangible Cultural Heritage}(2003)]%
        {unesco}
\bibfield{author}{\bibinfo{person}{{UNESCO Intangible Cultural Heritage}}.} \bibinfo{year}{2003}\natexlab{}.
\newblock \bibinfo{booktitle}{\emph{Intangible Cultural Heritage Domains}}.
\newblock UNESCO.
\newblock
\urldef\tempurl%
\url{https://ich.unesco.org/doc/src/01857-EN.pdf}
\showURL{%
\tempurl}


\bibitem[Vecco(2010)]%
        {definition_2010}
\bibfield{author}{\bibinfo{person}{Marilena Vecco}.} \bibinfo{year}{2010}\natexlab{}.
\newblock \showarticletitle{A definition of cultural heritage: From the tangible to the intangible}.
\newblock \bibinfo{journal}{\emph{Journal of Cultural Heritage}} \bibinfo{volume}{11}, \bibinfo{number}{3} (\bibinfo{year}{2010}), \bibinfo{pages}{321--324}.
\newblock
\showISSN{1296-2074}
\urldef\tempurl%
\url{https://doi.org/10.1016/j.culher.2010.01.006}
\showDOI{\tempurl}


\bibitem[Vosinakis et~al\mbox{.}(2018)]%
        {vosinakis2018dissemination}
\bibfield{author}{\bibinfo{person}{Spyros Vosinakis}, \bibinfo{person}{Nikos Avradinis}, {and} \bibinfo{person}{Panayiotis Koutsabasis}.} \bibinfo{year}{2018}\natexlab{}.
\newblock \showarticletitle{Dissemination of intangible cultural heritage using a multi-agent virtual world}. In \bibinfo{booktitle}{\emph{Advances in Digital Cultural Heritage: International Workshop, Funchal, Madeira, Portugal, June 28, 2017, Revised Selected Papers}}. Springer, \bibinfo{pages}{197--207}.
\newblock


\bibitem[Wang and Qin(2020)]%
        {lianying_art_2020}
\bibfield{author}{\bibinfo{person}{Lianying Wang} {and} \bibinfo{person}{Kuijie Qin}.} \bibinfo{year}{2020}\natexlab{}.
\newblock \bibinfo{booktitle}{\emph{Art of Traditional Chinese Flower Arrangement} (\bibinfo{edition}{first} ed.)}.
\newblock \bibinfo{publisher}{Chemical Industry Press}.
\newblock
\showISBNx{978-7-122-36279-7}


\bibitem[Wang and Zhang(2017)]%
        {wang2017aesthetics}
\bibfield{author}{\bibinfo{person}{Na Wang} {and} \bibinfo{person}{Yinghui Zhang}.} \bibinfo{year}{2017}\natexlab{}.
\newblock \showarticletitle{Aesthetics-based Beauty of Artistic Conception of Chinese Ancient Floral Arrangement}.
\newblock \bibinfo{journal}{\emph{Agricultural Science \& Technology}} \bibinfo{volume}{18}, \bibinfo{number}{9} (\bibinfo{year}{2017}), \bibinfo{pages}{1754--1756}.
\newblock


\bibitem[Wu et~al\mbox{.}(2023)]%
        {ArtPractice}
\bibfield{author}{\bibinfo{person}{Weilong Wu}, \bibinfo{person}{Jiaqing Lin}, {and} \bibinfo{person}{Zhiling Jiazeng}.} \bibinfo{year}{2023}\natexlab{}.
\newblock \showarticletitle{An Innovative Study on the Integration of Cross-Cultural Virtual Reality Technology in Art Practice Courses}. In \bibinfo{booktitle}{\emph{International Conference on Human-Computer Interaction}}. Springer, \bibinfo{pages}{61--72}.
\newblock


\bibitem[Xiong et~al\mbox{.}(2023)]%
        {OperARtistry}
\bibfield{author}{\bibinfo{person}{Zeyu Xiong}, \bibinfo{person}{Shihan Fu}, {and} \bibinfo{person}{Mingming Fan}.} \bibinfo{year}{2023}\natexlab{}.
\newblock \showarticletitle{OperARtistry: An AR-based Interactive Application to Assist the Learning of Chinese Traditional Opera (Xiqu) Makeup}. In \bibinfo{booktitle}{\emph{Proceedings of the Eleventh International Symposium of Chinese CHI}}. \bibinfo{pages}{158--168}.
\newblock


\bibitem[Xu(2023)]%
        {xu_inheritance_2023}
\bibfield{author}{\bibinfo{person}{Jie Xu}.} \bibinfo{year}{2023}\natexlab{}.
\newblock \showarticletitle{The Inheritance and Development of Traditional Chinese Flower Arrangement Art}.
\newblock \bibinfo{journal}{\emph{Highlights in Art and Design}} \bibinfo{volume}{4}, \bibinfo{number}{3} (\bibinfo{year}{2023}), \bibinfo{pages}{121--126}.
\newblock
\showISSN{2957-8787}
\urldef\tempurl%
\url{https://doi.org/10.54097/fpi2xlg1}
\showDOI{\tempurl}


\bibitem[Yang et~al\mbox{.}(2023)]%
        {yang2023impact}
\bibfield{author}{\bibinfo{person}{Yang Yang}, \bibinfo{person}{Zhengyun Wang}, \bibinfo{person}{Han Shen}, {and} \bibinfo{person}{Naipeng Jiang}.} \bibinfo{year}{2023}\natexlab{}.
\newblock \showarticletitle{The Impact of Emotional Experience on Tourists’ Cultural Identity and Behavior in the Cultural Heritage Tourism Context: An Empirical Study on Dunhuang Mogao Grottoes}.
\newblock \bibinfo{journal}{\emph{Sustainability}} \bibinfo{volume}{15}, \bibinfo{number}{11} (\bibinfo{year}{2023}), \bibinfo{pages}{8823}.
\newblock


\bibitem[Yoganathan et~al\mbox{.}(2018)]%
        {knottyingskills}
\bibfield{author}{\bibinfo{person}{Sutharsan Yoganathan}, \bibinfo{person}{David~A Finch}, \bibinfo{person}{E Parkin}, {and} \bibinfo{person}{J Pollard}.} \bibinfo{year}{2018}\natexlab{}.
\newblock \showarticletitle{360 virtual reality video for the acquisition of knot tying skills: A randomised controlled trial}.
\newblock \bibinfo{journal}{\emph{International Journal of Surgery}}  \bibinfo{volume}{54} (\bibinfo{year}{2018}), \bibinfo{pages}{24--27}.
\newblock


\bibitem[Yu et~al\mbox{.}(2024)]%
        {yujian}
\bibfield{author}{\bibinfo{person}{Jian Yu}, \bibinfo{person}{Zhan Wang}, \bibinfo{person}{Yifan Cao}, \bibinfo{person}{Hao Cui}, {and} \bibinfo{person}{Wei Zeng}.} \bibinfo{year}{2024}\natexlab{}.
\newblock \showarticletitle{Centennial Drama Reimagined: An Immersive Experience of Intangible Cultural Heritage through Contextual Storytelling in Virtual Reality}.
\newblock \bibinfo{journal}{\emph{ACM Journal on Computing and Cultural Heritage}} (\bibinfo{year}{2024}).
\newblock


\bibitem[Yu et~al\mbox{.}(2021)]%
        {guqin}
\bibfield{author}{\bibinfo{person}{Minjing Yu}, \bibinfo{person}{Meng Zhang}, \bibinfo{person}{Chun Yu}, \bibinfo{person}{Xiaoguang Ma}, \bibinfo{person}{Xing-Dong Yang}, {and} \bibinfo{person}{Jiawan Zhang}.} \bibinfo{year}{2021}\natexlab{}.
\newblock \showarticletitle{We can do more to save guqin: Design and evaluate interactive systems to make guqin more accessible to the general public}. In \bibinfo{booktitle}{\emph{Proceedings of the 2021 CHI Conference on Human Factors in Computing Systems}}. \bibinfo{publisher}{Association for Computing Machinery}, \bibinfo{address}{New York, NY, USA}, \bibinfo{pages}{1--12}.
\newblock


\bibitem[Zhang(2009)]%
        {ZhangJie}
\bibfield{author}{\bibinfo{person}{Jie Zhang}.} \bibinfo{year}{2009}\natexlab{}.
\newblock \emph{\bibinfo{title}{Preliminary Studies of Craft and Market for Industrialization of Chinese Traditional Flower Arrangement Art}}.
\newblock \bibinfo{thesistype}{Master's\ thesis}. \bibinfo{school}{Huazhong Agricultural University}, \bibinfo{address}{Wuhan, China}.
\newblock
\urldef\tempurl%
\url{https://kns.cnki.net/KCMS/detail/detail.aspx?dbname=CMFD2009&filename=2008202630.nh}
\showURL{%
\tempurl}


\bibitem[Zhang et~al\mbox{.}(2022)]%
        {touch}
\bibfield{author}{\bibinfo{person}{Yanxiang Zhang}, \bibinfo{person}{Yingna Wang}, {and} \bibinfo{person}{Qingqin Liu}.} \bibinfo{year}{2022}\natexlab{}.
\newblock \showarticletitle{Touch the History in Virtuality: Combine Passive Haptic with 360° Videos in History Learning}. In \bibinfo{booktitle}{\emph{2022 IEEE Conference on Virtual Reality and 3D User Interfaces Abstracts and Workshops (VRW)}}. \bibinfo{publisher}{IEEE}, \bibinfo{pages}{824--825}.
\newblock
\urldef\tempurl%
\url{https://doi.org/10.1109/VRW55335.2022.00263}
\showDOI{\tempurl}


\bibitem[Zhou et~al\mbox{.}(2024)]%
        {zhou2024integrating}
\bibfield{author}{\bibinfo{person}{Xuanxuan Zhou}, \bibinfo{person}{Qian Yang}, \bibinfo{person}{Linlin Bi}, {and} \bibinfo{person}{Siwang Wang}.} \bibinfo{year}{2024}\natexlab{}.
\newblock \showarticletitle{Integrating traditional apprenticeship and modern educational approaches in traditional Chinese medicine education}.
\newblock \bibinfo{journal}{\emph{Medical Teacher}} \bibinfo{volume}{46}, \bibinfo{number}{6} (\bibinfo{year}{2024}), \bibinfo{pages}{792--807}.
\newblock


\end{thebibliography}

% \clearpage

\appendix
\section{Appendix}

% \subsection{The questionnaires }

% \renewcommand{\thetable}{\arabic{table}} % 如果希望表格编号为 A1, A2 等
\setcounter{table}{0} % 如果需要重新开始编号

% \section{appendix}
% \captionsetup[table]{labelformat=empty} % 隐藏表格的 "Table" 前缀

\begin{table}[H]
  \caption{The questionnaires used in the One Session Practice User Study to assess FloraJing's system usability.}
  \Description{The questionnaires used in the One Session Practice User Study to assess FloraJing's system usability, includes  System Usability Scale and NASA-TLX. }
  \label{app:one-session-SUS-NASA}
  \renewcommand{\arraystretch}{1.4}
  \scalebox{0.72}{
   \begin{tabular}{ll}
\hline
\multicolumn{2}{c}{\textbf{I. System Usability Scale (SUS)}}                     \\ \hline
1  & I think that I would like to use FloraJing frequently.                      \\ \hline
2  & I found FloraJing unnecessarily complex.                                    \\ \hline
3  & I thought FloraJing was easy to use.                                        \\ \hline
4 & I think that I would need the support of a technical person to be able \\ & to use FloraJing. \\ \hline
5  & I found the various functions in FloraJing were well integrated.            \\ \hline
6  & I thought there was too much inconsistency in FloraJing.                    \\ \hline
7  & I imagine that most people would learn to use FloraJing very quickly.       \\ \hline
8  & I found FloraJing very awkward to use.                                      \\ \hline
9  & I felt very confident using FloraJing.                                      \\ \hline
10 & I needed to learn a lot of things before I could get going with FloraJing.  \\ \hline
\multicolumn{2}{c}{\textbf{II. NASA Task Load Index (NASA-TLX）}}                 \\ \hline
1  & How mentally demanding was it to complete the TCFA practice in FloraJing?   \\ \hline
2  & How physically demanding was it to complete the TCFA practice in FloraJing? \\ \hline
3  & How hurried or rushed was the pace of the TCFA practice in FloraJing?       \\ \hline
4  & How successful were you in accomplishing the TCFA practice in FloraJing?    \\ \hline
5  & How hard did you have to work to accomplish your level of performance?      \\ \hline
6  & How insecure, discouraged, irritated, stressed, and annoved were you?       \\ \hline
\end{tabular}
}
\end{table}

\begin{table}[hbp]
  \caption{The questionnaires used in the One Session Practice User Study to assess FloraJing's TCFA practice effect.}
  \Description{The questionnaires used in the One Session Practice User Study to assess FloraJing's TCFA practice effect, includes  Overall Practice Evaluation, Composition, Color Allocation, Philosophical Aesthetics, and Cultural Context. }
  \label{app:one-session-practice}
  \renewcommand{\arraystretch}{1.4}
  \scalebox{0.71}{
   \begin{tabular}{ll}
\hline
\multicolumn{2}{c}{\textbf{TCFA Practice Scale}}                                                      \\ \hline
\multicolumn{2}{c}{\textbf{I. Overall Practice Evaluation}}                                                   \\ \hline
1. & I feel that my overall skills in Chinese flower arrangement have improved.                      \\ \hline
\multicolumn{2}{c}{\textbf{II. Composition}}                                                          \\ \hline
2.   & I feel that my composition skills in Chinese flower arrangement have improved.                  \\ \hline
3.   & I feel that my shape-designing skills in Chinese flower arrangement have improved.              \\ \hline
4.   & I think that my proportion skills in Chinese flower arrangement have improved.                  \\ \hline
5.   & I think my work has improved in terms of balance and contrast.                                  \\ \hline
\multicolumn{2}{c}{\textbf{III. Color Allocation}}                                                    \\ \hline
6.   & I feel that my color allocation skills in Chinese flower arrangement have improved.             \\ \hline
7.   & I think my skill in the overall harmony of colors in my work has improved.                      \\ \hline
8.   & I think my skill to express the depth and variation of colors has improved.                     \\ \hline
\multicolumn{2}{c}{\textbf{IV. Philosophical Aesthetics}}                                             \\ \hline
9.   & I believe I have gained a deeper understanding of the philosophical aesthetics of TCFA          \\ \hline
10.  & I believe I have gained a deeper understanding of the Chinese philosophy of the unity  \\ & between nature and humanity.\\ \hline
11.  & I think I have a deeper understanding of the philosophical concept of Yin-Yang balance          \\ \hline
12. & I think my understanding of yijing has deepened.                                                \\ \hline
13. & I think I have a deeper understanding of the aesthetics of the pursuit of intended blank\\ & space. \\ \hline
\multicolumn{2}{c}{\textbf{V. Cultural Context}}                                                      \\ \hline
11. & I believe I have gained a deeper understanding of the cultural context of TCFA                  \\ \hline
12. & I think I have a deeper understanding of the symbolic meanings attributed to flowers \\ &in Chinese literature.        \\ \hline
13. & I have gained a deeper understanding of expressing emotions through flowers in \\ & Chinese culture. \\ \hline
\end{tabular}
}
\end{table}

\begin{table}[htbp]
  \caption{The outline of semi-structured interview according to Kolb's experiential earning model for the 7-day Pilot Longitudinal Study.}
  \Description{The semi-structured interview questions for the FloraJing VR User Study, organized according to Kolb's experiential learning model.}
  \label{app:7-day-interview}
  \footnotesize
  % \small% 缩小字号
  % \centering
  \renewcommand{\arraystretch}{1.4}
  \begin{tabularx}{\linewidth}{@{} l >{\RaggedRight}X @{}}
   
\hline
\multicolumn{2}{c}{\textbf{I. Concrete Experience}}             \\ \hline
1  & Can you describe your overall experience using the FloraJing VR system for TCFA? \\ \hline
2  & What specific features of FloraJing did you find most beneficial in enhancing your TCFA skills? \\ \hline
\multicolumn{2}{c}{\textbf{II. Reflective Observation}}             \\ \hline
3  & During your practice sessions, did you engage in any reflection about your skills or techniques? If so, how did this reflection manifest? \\ \hline
4  & What challenges did you encounter while practicing your TCFA skills in FloraJing, and how did you address them? \\ \hline
\multicolumn{2}{c}{\textbf{III. Abstract Conceptualization}}             \\ \hline
5  & In what ways did your VR practice deepen your understanding of TCFA culture and its significance? \\ \hline
6  & How did the interactive features of FloraJing contribute to your learning about floral design techniques and cultural practices? \\ \hline
\multicolumn{2}{c}{\textbf{IV. Active Experimentation}}             \\ \hline
7  & Have you practiced TCFA in real life after using FloraJing? If so, how did your VR experience influence your real-life floral arrangements?\\ \hline
8  & How do you plan to apply the skills you learned in FloraJing to real-life TCFA or other creative activities? \\ \hline
\multicolumn{2}{c}{\textbf{V. General Feedback}}             \\ \hline
9  & What aspects of the FloraJing most effectively supported your skill development? Were there any features you felt could be improved? \\ \hline
10 & Is there anything else you would like to share about your experience with FloraJing? \\ \hline
\end{tabularx}
\end{table}

\begin{table*}[htb]
    \centering
    \caption{\textbf{This table presents the demographics and relevant experience of participants, including prior experience with TCFA, and previous VR experience. \rv{$*$ indicates participants who took part in both User Study 1 and User Study 2.}}}
    \label{table:Demographic background}
    \Description{This table shows the demographics and relevant experience of 15 participants in our user study. ID, age, sex, prior experience with Traditional Chinese Flower Arrangement (TCFA), frequency of daily practice, and previous virtual reality experience are included in this table.}
    \resizebox{0.88\textwidth}{!}{
    \begin{tabular}{cccllllc}
   \hline
\multirow{2}{*}{\textbf{ID}} &
  \multirow{2}{*}{\textbf{Age}} &
  \multirow{2}{*}{\textbf{Sex}} &
  \multicolumn{3}{c}{\textbf{Prior TCFA Experience}} &
  \multirow{2}{*}{\textbf{\begin{tabular}[c]{@{}l@{}}Practice Frequency \end{tabular}}} &
  \multicolumn{1}{l}{\multirow{2}{*}{\textbf{\begin{tabular}[c]{@{}l@{}}Previous VR Experience\end{tabular}}}} \\ \cline{4-6}
    &    &   & Skill Level & Years of Experience & Total Time Spent       &                 & \multicolumn{1}{l}{} \\ \hline
P1$*$  & 28 & M & Beginner    & 4 months         & 5 hours                & Every 2 months  & Yes                  \\
P2  & 26 & F & Beginner    & 3 months         & 10 hours               & Every 2 weeks   & Yes                  \\
P3  & 23 & M & Beginner    & 1 month          & \textless 5 hours      & Every month     & Yes                  \\
P4  & 22 & F & Beginner    & 1 year           & 15 hours               & Every 2 months  & Yes                  \\
P5  & 20 & F & Beginner    & 3 months         & 5 hours                & Every 2 months  & No                   \\
P6$*$  & 42 & F & Beginner    & 2 months         & \textless 5 hours      & Every month     & Yes                  \\
P7  & 25 & M & Competent   & 3 months         & 40 hours               & Every week      & Yes                  \\
P8  & 28 & M & Competent   & 6 months         & 50 hours               & Every 2 weeks   & Yes                  \\
P9  & 20 & M & Competent   & 8 months         & 50 hours               & Every week      & No                   \\
P10$*$ & 26 & F & Competent   & 2 years          & 70 hours               & Every month     & Yes                  \\
P11$*$ & 37 & F & Proficient  & 10 years         & 200 hours              & Every 2 months  & No                   \\
P12$*$ & 28 & F & Proficient  & 2 years          & 150 hours              & Every two weeks & No                   \\
P13 & 20 & M & Proficient  & 1 year           & 120 hours              & Every month     & Yes                  \\
\rv{P14} & \rv{38} & \rv{M} & \rv{Proficient}  & \rv{2.5 year}           & \rv{150 hours}              & \rv{Every month}     & \rv{No}                 \\
P15$*$ & 26 & F & Expert      & 10 years         & \textgreater 200 hours & Every week      & Yes                  \\
P16$*$ & 37 & M & Expert      & 5 years          & \textgreater 200 hours & Every 2 day     & No                   \\
\rv{P17} & \rv{33} & \rv{M} & \rv{Expert}      & \rv{4 years}          & \textgreater \rv{200 hours} & \rv{Every 2 day}     & \rv{No}                   \\
\rv{P18} & \rv{41} & \rv{F} & \rv{Expert}      & \rv{7 years}          & \rv{\textgreater 200 hours} & \rv{Every week}     & \rv{No}                   \\ 
\rv{P19} & \rv{39} & \rv{M} & \rv{Expert}      & \rv{9 years}          & \rv{\textgreater 200 hours} & \rv{Every week}     & \rv{Yes}                   \\  \hline
\end{tabular}
}
\end{table*}


\end{document}